\documentclass[aps,pra,superscriptaddress,longbibiliography]{revtex4-2}

\usepackage{graphicx}
\usepackage{xcolor}
\usepackage{braket}

\usepackage{bm}
\usepackage[colorlinks=true,breaklinks=true,linkcolor=red,citecolor=magenta,urlcolor=blue]{hyperref}
\usepackage{comment}

\usepackage[utf8]{inputenc}
\usepackage[T1]{fontenc}
\usepackage{qcircuit}
\usepackage{float}
\usepackage{amsmath,amsfonts,amsthm}

\newcommand{\numberop}[1]{\hat{n}_{#1}}
\newcommand{\vett}[1]{{\boldsymbol{#1}}}
\newcommand{\crt}[1]{ \hat{a}^\dagger_{#1} }
\newcommand{\dst}[1]{ \hat{a}^{\phantom{\dagger}}_{#1} }

\newcommand{\reference}{ \Phi_0 }

\begin{document}

\title{Quantum algorithms for the variational optimization of correlated electronic states with stochastic reconfiguration and the linear method}

\author{Mario Motta}
\email{mario.motta@ibm.com}
\affiliation{IBM Quantum, IBM T.J. Watson Research Center, Yorktown Heights, NY 10598, USA}
\author{Kevin J. Sung}
\affiliation{IBM Quantum, IBM T.J. Watson Research Center, Yorktown Heights, NY 10598, USA}
\author{James Shee$^*$}
\email{james.shee@rice.edu}
\affiliation{Department of Chemistry, Rice University, Houston, TX 77005, USA}

\begin{abstract}
\noindent Solving the electronic Schrodinger equation for strongly correlated ground states is a long-standing challenge. We present quantum algorithms for the variational optimization of wavefunctions correlated by products of unitary operators, such as Local Unitary Cluster Jastrow (LUCJ) ansatzes, using stochastic reconfiguration and the linear method.  While an implementation on classical computing hardware would require exponentially growing compute cost, the cost (number of circuits and shots) of our quantum algorithms is polynomial in system size.  We find that classical simulations of optimization with the linear method consistently find lower energy solutions than with the L-BFGS-B optimizer across the dissociation curves of the notoriously difficult N$_2$ and C$_2$ dimers; LUCJ predictions of the ground-state energies deviate from exact diagonalization by 1 kcal/mol or less at all points on the potential energy curve.  While we do characterize the effect of shot noise on the LM optimization, these noiseless results highlight the critical but often overlooked role that optimization techniques must play in attacking the electronic structure problem (on both classical and quantum hardware), for which even mean-field optimization is formally NP hard. We also discuss the challenge of obtaining smooth curves in these strongly correlated regimes, and propose a number of quantum-friendly solutions ranging from symmetry-projected ansatz forms to a symmetry-constrained optimization algorithm. 
\end{abstract}

\maketitle

\section{Introduction}

The emergence and rapid improvement of quantum computing hardware has opened new research directions in the molecular electronic structure community.  With a focus on challenging regimes that formally involve computational cost scaling exponentially with system size on classical computers -- typically involving themes of strong correlation and real-time propagation -- the research community has proposed alternative and complementary algorithms using quantum computing platforms to solve such problems.  In this work, we present quantum algorithms that provide a compelling alternative to a stochastic approach to wavefunction parameter optimization on classical computers, namely, variational Monte Carlo (VMC).  Our algorithms enables optimization with stochastic reconfiguration (SR) or the linear method (LM) such that the quantum device is used to evaluate classically-intractable matrix elements.  Classical simulations of these optimizers are shown to converge to approximate ground-states 
of molecular Hamiltonians of notoriously challenging dissociation coordinates in a manner that is more robust and efficient than widely-used quasi-Newton optimizers.

A leading strategy to use near-term devices for quantum chemistry is the Variational Quantum Eigensolver (VQE)~\cite{peruzzo2014variational}.  This scheme computes the energy of a parameterized wavefunction ansatz on the quantum device and optimizes the variational parameters classically until convergence. While in practice only relatively small electronic systems have been studied, due to hardware limitations, VQE has the potential to variationally optimize correlated wavefunction forms that are size-extensive and unitary -- currently such a task would incur an exponentially growing classical computational cost as a function of system size.  Examples of such wavefunctions, typically with non-linear correlators, include trotterized Unitary Coupled Cluster Singles and Doubles (qUCCSD)~\cite{anand2022quantum} and the Unitary Cluster Jastrow (UCJ)~\cite{matsuzawa2020jastrow}.  An alternative strategy involves ansatzes that are, by construction, hardware-efficient; examples include qubit coupled cluster\cite{ryabinkin2018qubit,ryabinkin2020iterative} and adaptive wavefunction forms\cite{grimsley2019adaptive,tang2021qubit}.

Recently, we have presented a family of local approximations to the UCJ that is both physically motivated and quantum hardware-friendly (LUCJ)~\cite{LUCJ}.  We found that the LUCJ ansatz, which retains only on-site opposite-spin and nearest-neighbor same-spin interactions, could achieve higher accuracy than, e.g., qUCCSD with shallower circuits.  However, as the number of wavefunction parameters increases and/or in regimes of strong electron correlation,~\cite{ganoe2024notion}  we observed that optimization with the quasi-Newton method, L-BFGS-B~\cite{byrd1995lbfgsb,zhu1997algorithm}, at times required a large number of iterations (c.f. Fig. 14 of Ref. \citenum{LUCJ}).  Furthermore, such optimization is likely susceptible to converging on local minima (indeed, even the optimization of mean-field wavefunctions is formally NP hard\cite{chan2024quantum}).
The search for a more robust optimizer for ansatzes like LUCJ is one motivation for the present work. 

Fortunately, the problem of variationally optimizing a large number of parameters, indeed even in Jastrow-type wavefunctions, has been thought about for decades in the context of classical computing in the VMC community~\cite{foulkes2001quantum}.  Orbital-space VMC is arguably most natural in the context of quantum computing, and much progress has been made in classical algorithmic design~\cite{wei2018reduced,sabzevari2018improved,mahajan2020efficient}.   Indeed, there is a rather established hierarchy of quasi-second-order optimization methods ranging from SR~\cite{SR_Sorella} to the LM~\cite{nightingale2001optimization,toulouse2007optimization,toulouse2008full,Umrigar_2007}.  
These are more effective than optimizers based on the energy gradient alone since the involved equations involve not only energy gradients but also first and second derivatives of the wavefunction with respect to its variational parameters.  Indeed, when the wavefunction can be exactly expanded to linear order, the LM will be equivalent to a full second-order optimizer, e.g., the Newton method~\cite{UmrigarFilippi2005}.  We note that the relative performance of methods such as the LM vs those which approximate the inverse of the energy Hessian, such as L-BFGS, is not obvious, and we seek to answer this question empirically.  
In what follows, we show numerical results from classical simulations of 
LM-optimized LUCJ ansatzes, and present quantum ``subroutines'' for the variational optimization of LUCJ wavefunctions via SR and the LM. 
Importantly, our quantum SR (qSR) and LM (qLM) algorithms permit the optimization of wavefunction forms that are infeasible to optimize classically, even with VMC techniques.  

\section{Stochastic Reconfiguration and Linear Method}

\subsection{SR}
Our objective is to optimize a normalized orbital-space wavefunction $\Psi(\vett{\theta})$ with $N_\theta$ variational parameters $\vett{\theta}$ by minimizing the energy $E(\vett{\theta}) = \langle \Psi(\vett{\theta}) | \hat{H} | \Psi(\vett{\theta}) \rangle$. 
Evolving $\Psi$ in imaginary time via a first-order propagator expansion gives $e^{-\tau \hat{H}} | \Psi \rangle \rightarrow 1-\tau\hat{H} | \Psi \rangle$.  Projecting onto the space spanned by the current wavefunction, $|\Psi^0\rangle = |\Psi\rangle$, and its first derivatives,  $|\Psi^i \rangle = \frac{\partial}{\partial \theta_i} |\Psi \rangle $, yields a wavefunction of the form $|\Psi' \rangle = \sum_{i=0}^{N_\theta} x_i |\Psi^i \rangle$, where the $x$ values are defined from
\begin{equation}
    \langle \Psi^j | (1 - \tau \hat{H})|\Psi\rangle = \sum_i \langle \Psi^j | \Psi^i \rangle x_i.
    \label{SRlinear}
\end{equation}
By eliminating the factor $x_0$ through the $\mu=0$ equation, one derives a set of equations for $x_1 \dots x_{N_\theta}$, which can be written in matrix notation as
\begin{equation}
\label{eq:lin_eq_1}
    \mathbf{S} \mathbf{x} = -\frac{\tau}{2} \mathbf{g}
    \;,
\end{equation}
where the derivative overlap matrix is defined as 
\begin{equation}
S_{ij} = \mbox{Re}\left[ \langle \Psi^i | \Psi^j \rangle - \langle \Psi^i | \Psi \rangle \langle \Psi | \Psi^j \rangle \right]
\label{eq:overlap}
\end{equation}
and the gradient vector by 
\begin{equation}
    g_i = 2 \, \mbox{Re}[ \langle \Psi^i | \hat{H} | \Psi \rangle - E(\vett{\theta})  \langle \Psi^i | \Psi \rangle ]
    = \frac{\partial E}{\partial \theta_i}(\vett{\theta})
    \;.
    \label{eq:gradient}
\end{equation}
The updated wavefunction parameters are then computed via $\theta^\prime_i = \theta_i + x_i$.

\subsection{LM}

The linear method takes a normalized wavefunction, 
$\Psi$, and expands it to first-order in the variational parameters,
\begin{equation}
    |\Psi'\rangle = | \Psi^0 \rangle + \sum_{i=1}^{N_\theta} x_i | \Psi^i \rangle \;,
\end{equation}
where $|\Psi^0 \rangle = |\Psi\rangle$, $| \Psi^i \rangle = | \Psi^i \rangle - | \Psi^0 \rangle \langle \Psi^0 | \Psi^i \rangle$, and $x_0$ is assumed here to be one.

The variational energy is minimized by the condition
$\partial_{\vett{x}} \langle \Psi^\prime | \hat{H} | \Psi^\prime \rangle = 0$,
yielding the generalized eigenvalue equation
\begin{equation}
\begin{pmatrix}
E(\vett{\theta}) & \frac{1}{2}\mathbf{g}^T \\
\frac{1}{2} \mathbf{g} & \mathcal{H}  
\end{pmatrix}
\begin{pmatrix}
1 \\
\boldsymbol{x}
\end{pmatrix}
=
E_0
\begin{pmatrix}
1 & 0 \\
0 & \mathcal{S}
\end{pmatrix}
\begin{pmatrix}
1 \\
\boldsymbol{x}
\end{pmatrix}
\;,
\label{eq:hamiltonian}
\end{equation}
where $\boldsymbol{g}$ is the gradient expression in \eqref{eq:gradient}, 
$\mathcal{H}_{ij} = \mbox{Re}[\langle \Psi^i | \hat{H} | \Psi^j \rangle]$, and
$\mathcal{S}_{ij}$ is defined in Eq.~\eqref{eq:overlap}.
As in SR, the parameter update is given by $\theta'_i = \theta_i + x_i$. 

When the wavefunction at a given iteration is outside the quadratic convergence basin and/or the ``Hessian,'' $\mathcal{H}$, is not well-resolved (e.g., due to insufficient MC sampling or quantum shot noise), the diagonals of $\mathcal{H}$ can be shifted by a real, positive quantity $\alpha$ to restore positive-definiteness, and more generally to damp the magnitude of parameter changes.  This shift also rotates the update vector from the Newton-like direction to that from steepest-descent.  This Tikhonov regularization scheme was first presented in a seminal work in VMC optimization\cite{UmrigarFilippi2005}.  
The LM equations can be recast as a Newton method,
\begin{equation}
\label{eq:lin_eq_2}
    (\mathbf{A} + \alpha \mathbf{S}) \mathbf{x} = -\frac{1}{2}\mathbf{g}
    \;,
\end{equation}
where $\mathbf{A} = \mathbf{H} - E(\vett{\theta}) \mathbf{S}$ is an approximate and level-shifted Hessian, and $\alpha = E(\vett{\theta}) - E_0 > 0$.

Finally, we note that another parameter, $\xi$ $\in$ [0,1],  can be introduced to fine-tune the magnitude of parameter updates, such that
\begin{equation}
\label{eq:rescaling_1}
    \boldsymbol{x}' = \frac{\boldsymbol{x}}{1 + \frac{(1-\xi)Q}{(1-\xi)+\xi\sqrt{1+Q}}}
\end{equation}
with $Q = \sum_{jk} x_j S_{jk} x_k$.  Further details about our incorporation of the hyper-parameters $\alpha$ and $\xi$ will be given in Section III.A.1. 

\subsection{Motivation for a Quantum Subroutine}

Classical computation of matrices of the form
\begin{equation}
B_{ij} = \langle \partial_i \Psi | \hat{B} | \partial_j \Psi \rangle,
\label{eq:deriv}
\end{equation}
where $\hat{B}=\hat{I}$ or $\hat{H}$ for the overlap or Hamiltonian matrix, respectively, requires VMC. However, while specific types of wavefunction ansatzes -- such a Slater determinant (or a linear combination of multiple Slater determinants) correlated by a real-space Jastrow factor -- can be used in efficient classical VMC schemes, other types -- such as qUCCSD and (L)UCJ -- cannot, since the exact calculation of $|\langle \vett{x} | \Psi (\vett{\theta})\rangle|^2 = |\Psi_{\vett{\theta}}(\vett{x})|^2$ (required to sample the variational energy estimate) is not known to be tractable on classical computers.  

In what follows we first introduce the LUCJ ansatz and show numerical results of classically-simulated LM optimization  for the dissociation of two notoriously difficult diatomic dissociations, the N$_2$ and C$_2$ dimers.  We then derive quantum circuits that compute the expectation values involving wavefunction derivatives as required by SR and the LM in the context of the LUCJ ansatz, and analyze the corresponding measurement scaling.

\section{LUCJ Ansatz}

The LUCJ ansatz ~\cite{LUCJ,matsuzawa2020jastrow,motta2021low} is a product of exponentials of density-density operators interspersed with orbital rotations,
\begin{equation}
| \Psi(\vett{\theta}) \rangle = e^{-K_L} \prod_{\mu=0}^{L-1} e^{\hat{K}_\mu} e^{i \hat{J}_\mu} e^{-\hat{K}_\mu} | \reference \rangle \;,
\end{equation}
where
\begin{equation}
\begin{split}
\hat{K}_\mu &= \sum_{p>r} \kappa^{\mu 1}_{pr} \left[ \sum_\sigma \crt{p\sigma} \dst{r\sigma} - \crt{r\sigma} \dst{p\sigma} \right] + \sum_{p \geq r} \kappa^{\mu 2}_{pr} \left[ \sum_\sigma i \crt{p\sigma} \dst{r\sigma} + i \crt{r\sigma} \dst{p\sigma} \right] \;, \\
\hat{J}_\mu &= \sum_{p \in S} J^{\mu 1}_{p} \numberop{p\uparrow} \numberop{p\downarrow} + \sum_{pr \in S^\prime} J^{\mu 2}_{pr} \left[ \sum_\sigma \numberop{p\sigma} \numberop{r\sigma} \right] \;.
\end{split}
\end{equation}
The LUCJ Ansatz has $N_\theta = (L+1) N^2 + L (|S|+|S^\prime|)$ parameters, with $|S|,|S^\prime|=O(N)$ in typical implementations. 

Furthermore, in the standard Jordan-Wigner representation, it is implemented by a quantum circuit~\cite{LUCJ} of the form 
\begin{equation}
| \Psi(\vett{\alpha}) \rangle = \prod_{g=1}^{N_g} \hat{U}_g(\alpha_g) | \reference \rangle \;,
\end{equation}
where
\begin{enumerate}
\item the gates $U_g$ are either $\mathsf{XX+YY}$ gates,
\begin{equation}
\label{eq:xxyy_gate}
U_{\mathsf{XX+YY}}(\alpha) = 
\left( \begin{array}{crrc} 
1 & 0 & 0 & 0 \\
0 & \cos(\alpha) & -i \sin(\alpha) & 0 \\
0 & i\sin(\alpha) & \cos(\alpha) & 0 \\
0 & 0 & 0 & 1 \\
\end{array}
\right)
\;,
\end{equation}
phase gates ($\mathsf{P}$ gates),
\begin{equation}
\label{eq:p_gate}
P(\alpha) = 
\left( \begin{array}{cc} 
1 & 0 \\
0 & e^{i\alpha} 
\end{array}
\right)
\;,
\end{equation}
or density-density gates ($\mathsf{nn}$ gates)
\begin{equation}
\label{eq:nn_gate}
U_{\mathsf{nn}}(\alpha) = 
\left( \begin{array}{cccr} 
1 & 0 & 0 & 0 \\
0 & 1 & 0 & 0 \\
0 & 0 & 1 & 0 \\
0 & 0 & 0 & e^{i\alpha} \\
\end{array}
\right)
\;.
\end{equation}
\item the angles ${\boldsymbol{\alpha}}(\vett{\theta})$ are functions of the parameters $\vett{\theta} = (K_L , J_{L-1},K_{L-1}, \dots J_0,K_0).$
\item the total number of gates (and angles) is 
\begin{equation}
N_g = (L+1) 2 N^2 + L \big( |S|+2 |S^\prime| \big) \;.
\end{equation}
\end{enumerate}
Table \ref{tab:cost} lists the number of one- and two-qubit (cZ) gates for the circuits involved in our studies of the N$_2$ dissociation. Assuming a representative error rate of $\varepsilon_{\mbox{2q}} = 0.1 \%$ for 2-qubit gates yields the following rough estimates for the fidelity between the ideal and noisy circuit output state, $F = (1-\varepsilon_{\mbox{2qg}})^{N_{2q}} \simeq 0.853$ to $0.453$.  This estimate does \emph{not} account for any form of readout/gate error mitigation, which in the near term ought to be possible \cite{kandala2019error,czarnik2021error,nation2021scalable}.

We will adopt the shorthand notations $| \Psi(\vett{\alpha}) \rangle = \prod_{g=0}^{G-1} U_g | \reference \rangle$, $U^b_a = \prod_{g=a}^b U_g$, and 
$| \Phi_g \rangle = U_0^g | \reference \rangle$. 

\begin{table}
\begin{tabular}{cccc}
\hline\hline
 $L$ & $N_{1q}$ & $N_{2q}$ & Depth \\
\hline
 2 &  696             &  372         &     153 \\
 4 & 1256             &  684         &     281 \\
 6 & 1816             &  996         &     409 \\
 8 & 2376             & 1308         &     537 \\
10 & 2936             & 1620         &     665 \\
12 & 3496             & 1932         &     793 \\
\hline\hline
\end{tabular}
\caption{
Number of layers $L$, 1-qubit gates $N_{1q}$, cZ gates $N_{2q}$, and depth of the LUCJ quantum circuit for the N$_2$ molecule studied in this work.  In all cases, the number of qubits is $16$ and square-lattice connectivity is used.  }
\label{tab:cost}
\end{table}

\subsection{Numerical Results via Classical Simulation}
\subsubsection{Computational Details}
All calculations were performed using open-source software libraries.
We use PySCF~\cite{sun2018pyscf,sun2020recent} to generate the molecular integrals and
ffsim~\cite{ffsim} to simulate the quantum circuits,
and we use the implementation of L-BFGS-B from Scipy~\cite{scipy}.
Throughout, we use minimal STO-6G basis sets.

In our implementation of the linear method, we allow the hyperparameters $\alpha$ and $\xi$ to change with each iteration, rather than using fixed values throughout. In each iteration, we choose values of $\alpha$ and $\xi$ that minimize the energy of the resulting state. To choose the values, we run a minimization using L-BFGS-B, initialized with the values of $\alpha$ and $\xi$ used in the previous iteration. For the first iteration, we set $\alpha = 0$ and $\xi = 0.5$. After collecting the data for this work, we found that using a small but nonzero initial value for $\alpha$ seemed to be more robust in the absence of bootstrapping, so in practice, we recommend initializing the $\alpha$ parameter as $10^{-4}$. We have contributed our implementation to ffsim~\cite{ffsim}.

We employ LUCJ ansatzes that correspond to square qubit connectivity, with restricted Hartree-Fock (RHF) references.  We do not require that the optimized wavefunction be an eigenstate of spin nor spatial symmetries, and show the $\langle S^2 \rangle$ expectation values -- which typically do not deviate substantially from spin-pure quantities -- in the Appendix. 

For N$_2$ and C$_2$ dissociation curves we use active space sizes of 10 electrons in 8 orbitals and 8 electrons in 8 orbitals, respectively.  These correlated subspaces are formed from RHF molecular orbitals. We separately permuted the occupied and virtual RHF molecular orbitals to ensure they evolve continuously with bondlength.  We stop the LM optimization procedure when the gradient norm falls below a threshold of $10^{-5}$ in the noiseless simulations.  

The variational optimization of approximate strongly correlated electronic states is complicated by the presence of multiple nearly-degenerate eigenstates.  N$_2$~\cite{fan2006usefulness,bulik2015can} and especially C$_2$~\cite{abrams2004full,purwanto2009excited,booth2011breaking} represent prototypical challenges for optimization algorithms due to their many possible local minima.  It is well-appreciated that the initial guess of parameters can affect the cost and robustness of optimization algorithms, and we choose to initialize the LUCJ $K$ and $J$ parameters from those implied by a nested singlular value decomposition of the $t_2$ operator from qUCCSD.  
We then ``bootstrap'' -- i.e., initialize a calculation at $R_i=R_{i-1} + 1$ with converged parameters from the $R_{i-1}$ bond length -- all parameters \emph{except} for those in the $\exp(\hat{K}_L)$ (left-most) orbital rotation operator (at this stage omitted from the optimization, and set to zero), going from left to right, right to left, and then left to right again along the dissociation coordinate.  Among these, we take the lowest-energy solution and then restart an optimization that includes the left-most orbital rotation operator, initialized to the identity at every $R$.  Admittedly this procedure is not optimal, but for the purpose of this paper was established to enable a fair comparison between L-BFGS-B and the LM. 

\subsubsection{Nitrogen and carbon dimers}

\begin{figure}[h!]
\includegraphics[width=0.7\textwidth]{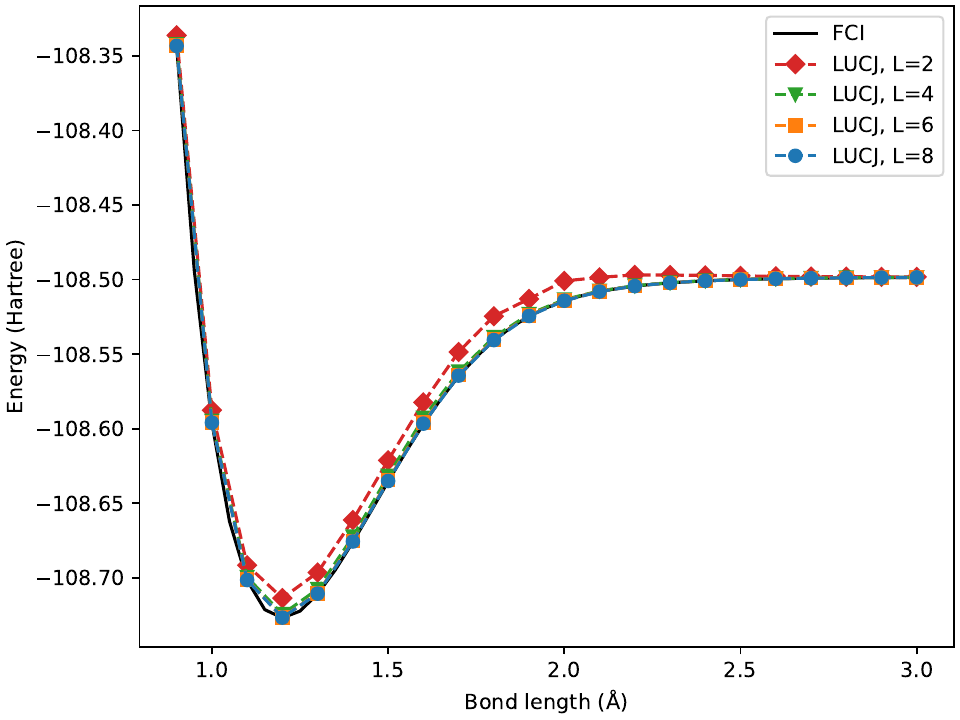}
\caption{N$_2$ dissociation in an (8e,10o) active space.  Classical simulation of the LM-optimized LUCJ ansatz with various values of $L$.}
\label{fig:N2energy}
\end{figure}

We begin our investigation with the dissociation of dinitrogen, which exhibits a triple bond at its equilibrium bond length.  Fig. \ref{fig:N2energy} compares the total energies of the LM-optimized LUCJ ansatzes for $L$ values of 2, 4, 6, and 8 vs the exact Full Configuration Interaction (FCI) energy in the same basis set.  As expected, given the increase in the number of variational parameters in the LUCJ ansatz, the energy decreases monotonically at each bond length as the number of layers in the LUCJ correlator, $L$, is increased.  It is noteworthy that, unlike related methods such as the unitary version of pair coupled cluster with double excitations~\cite{anand2022quantum}, LUCJ with these small values of $L$ is size-consistent and qualitatively correct even throughout the challenging regime at intermediate bond lengths.

\begin{figure}[h!]
\includegraphics[width=0.7\textwidth]{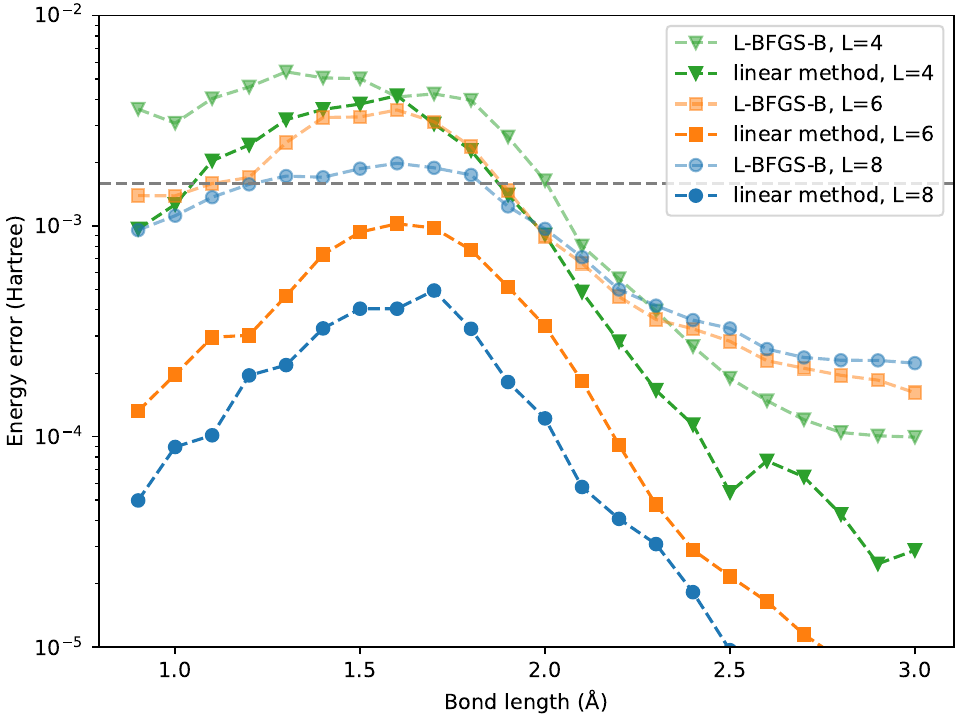}
\caption{N$_2$ dissociation energy errors vs FCI, comparing L-BFGS-B and LM optimizers for LUCJ ansatzes with $L$=4, 6, and 8.}
\label{fig:N2error}
\end{figure}

Fig. \ref{fig:N2error} zooms in on the energy errors vs FCI, focusing on $L$ values of 4, 6, and 8.  At this energy scale, a meaningful comparison between the L-BFGS-B and LM optimizers can be made (with both using the bootstrapping protocol described in the previous section).   
With the exception of only one bond length at $L=4$, the LM optimizer finds LUCJ solutions that are lower in energy across the board.  The $L$=6 and 8 results are particularly stark, with the LM-optimized LUCJ energy lower than the L-BFGS-B analog by an order of magnitude in multiple regions of the potential energy surface.  Indeed, the LM-optimized LUCJ with $L=6$ is already within the target range, i.e. sub-1.6 milliHartrees (1 kcal/mol) vs the exact diagonalization energy. 
 We note small yet apparent discontinuities in some of the curves, but have overlooked these given the relatively small scale of energy fluctuations.  

\begin{figure}[h!]
\includegraphics[width=0.7\textwidth]{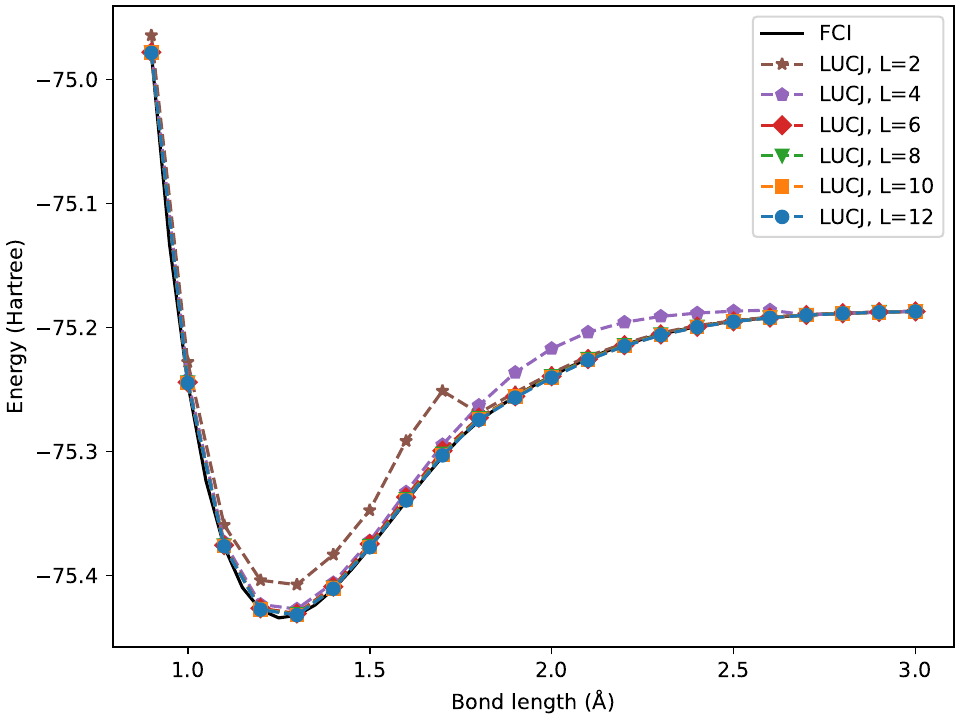}
\caption{C$_2$ dissociation in a (8e,8o) active space.  Classical simulation of the LM-optimized LUCJ ansatz with various values of $L$.}
\label{fig:C2energy}
\end{figure}

\begin{figure}[h!]
\includegraphics[width=0.7\textwidth]{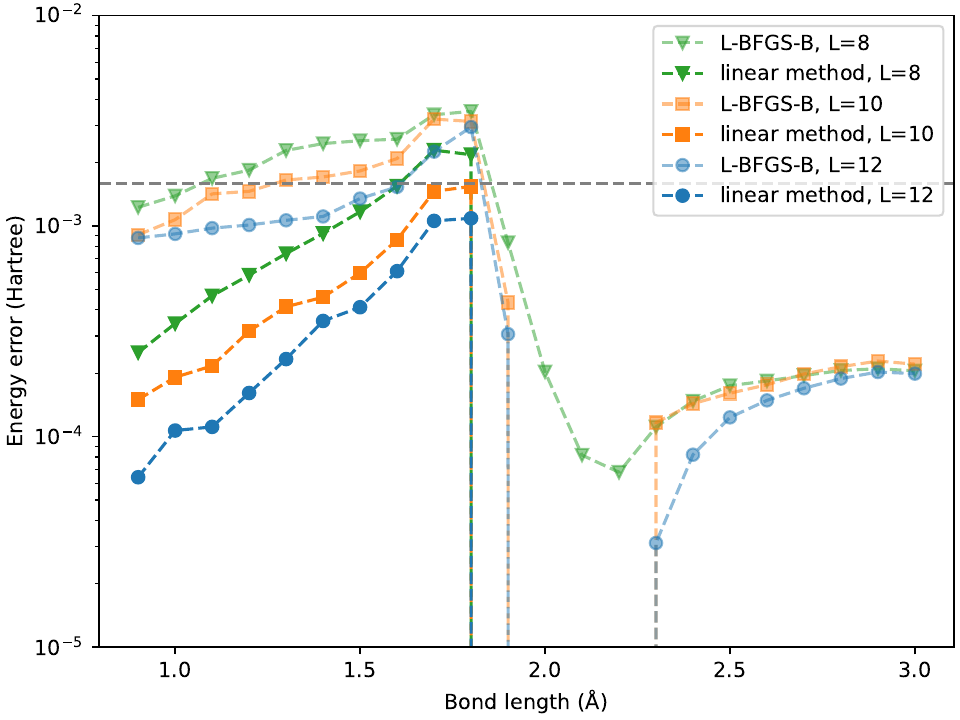}
\caption{C$_2$ dissociation energy errors vs FCI, comparing L-BFGS-B and LM optimizers for LUCJ ansatzes with $L$=8, 10, and 12.}
\label{fig:C2error}
\end{figure}

The dissociation energy curves of the carbon dimer are shown in Fig. \ref{fig:C2energy}.  Around the equilibrium bond length we observe the expected ordering of the curves corresponding to different $L$ values; however, to the right of roughly 1.8\AA \ the $L$=4 curve is higher in energy than the $L$=2 curve.  This highlights that the bootstrapping protocol that we consistently employ is not guaranteed to always find global minima (possible solutions will be discussed in the final section).  
Fig. \ref{fig:C2error} shows the energy errors of the various LUCJ ansatzes vs FCI.  Again we find that for fixed $L$ the LM-optimized LUCJ energy is consistently lower than that optimized with L-BFGS-B at every bond length.  Remarkably, the $L$=10 (and 12) LM-optimized LUCJ wavefunction is within 1.6 milliHartrees from the exact energy at all bond lengths. 

Taken together, we find that in these strongly correlated systems the widely-used L-BFGS-B algorithm frequently converges to local minima.  The LM is clearly to be preferred in terms of accuracy, and in the next section we introduce a quantum algorithm to compute the required quantities for optimization via the LM (with trivial extension to SR).

\pagebreak
\newpage

$ $

\pagebreak
\newpage

\section{Quantum algorithms}

To implement the LM and SR algorithms, it is necessary to evaluate the energy gradient in Eq.~\eqref{eq:gradient} along with the overlap matrix in Eq.~\eqref{eq:overlap} and the Hamiltonian matrix in Eq.~\eqref{eq:hamiltonian}.

Before estimating the number of measurements required to compute these quantities, let us recall that the Hamiltonian can be measured with a linearly scaling number of circuits, as shown by Huggins et al ~\cite{huggins2021efficient} (other efficient measurement schemes are described in, e.g., Ref.s \citenum{choi2023fluid,choi2022improving,yen2023deterministic,jena2022optimization,huang2021efficient}). One can use a low-rank decomposition of the electron repulsion integral to write
\begin{equation}
\hat{H} 
= \hat{h} + \frac{1}{2} \sum_{\gamma=1}^{N_\gamma} \hat{X}^2_\gamma
= \hat{U}_0^\dagger \hat{N}_0 \hat{U}_0
+ \frac{1}{2} 
\sum_{\gamma=1}^{N_\gamma}
\hat{U}_\gamma^\dagger \hat{N}_\gamma^2 \hat{U}_\gamma
\;,\;
\hat{N}_0 = \sum_{p\sigma} \eta_p \numberop{p\sigma}
\;,\;
\hat{N}_\gamma = \sum_{p\sigma} \xi^\gamma_p \numberop{p\sigma}
\end{equation}
where $N_\gamma = O(N)$~\cite{beebe1977simplifications,purwanto2011assessing} and $\hat{h}_1, \, \hat{X}_\gamma$ one-body operators diagonalized by Bogoliubov transformations $\hat{U}_0, \, \hat{U}_\gamma$. 
The expectation value $\mbox{Tr}[\hat{H} \hat{\rho}]$ can be estimated preparing the $(1+N_\gamma)$ states $\hat{U}_\gamma \hat{\rho} \hat{U}_\gamma^\dagger$ and measuring $\hat{N}_0, \, \hat{N}_\gamma$ using $S_0, \, S_\gamma$ statistical samples or ``shots''. The resulting variance is bounded by
\begin{equation}
\label{eq:energy_variance_0}
\mbox{Var}[\hat{H}] \leq \frac{4 \| \eta \|^2_1}{S_0} + \sum_\gamma \frac{4 \| \xi^\gamma \|_1^4}{S_\gamma}
\;,
\end{equation}
The optimal (i.e. lowest-variance) allocation of $S = S_0 + \sum_\gamma S_\gamma$ shots into $(1+N_\gamma)$ groups is found~\cite{rubin2018application} using the Lagrange multiplier technique: one minimizes the quantity $V = \sum_k w_k^2 / S_k$ under the constraint $\sum_k S_k = S$, finding $S_k = |w_k| S / \| \vett{x} \|_1$ and $V = \| \vett{w} \|_1^2 /S$. In the case of Eq.~\eqref{eq:energy_variance_0}, the Lagrange multiplier technique yields
\begin{equation}
\label{eq:energy_variance}
\mbox{Var}[\hat{H}] \leq \frac{\Lambda^2}{S} 
\;,\;
\Lambda = 2 \| \eta \|_1 + \sum_\gamma 2 \| \xi^\gamma \|_1^2
\;.
\end{equation}
The quantity $\Lambda$ scales between $O(N)$ and $O(N^3)$ depending on details of the particular system \cite{motta2019efficient,berry2019qubitization,lee2021even,von2021quantum}.

\subsection{Energy gradient}

The energy gradient in Eq.~\eqref{eq:gradient} can be computed applying the formulas called ``shift rules'' to the LUCJ circuit~\cite{wierichs2022general,hubregtsen2022single,izmaylov2021analytic}. For this purpose, it will be useful to observe that the gates Eq.~\eqref{eq:xxyy_gate}, \eqref{eq:p_gate}, and \eqref{eq:nn_gate} can be written as 
$\hat{U}_g(\alpha_g) = u_{00} + u_{10} \cos(\alpha_g) + u_{01} \sin(\alpha_g)$.
Therefore, the energy (as well as the expectation value of any operator independent of $\alpha_g$) depends on $\alpha_g$ (with all other angles fixed) as a second-degree trigonometric polynomial,
\begin{equation}
\begin{split}
E(\alpha_g) &= \langle \Phi_g | \hat{U}_g(\alpha_g)^\dagger \left( U_{g+1}^{N_g-1} \right)^\dagger \hat{H} U_{g+1}^{N_g-1} \hat{U}_g(\alpha_g) | \Phi_g \rangle \\
&=
\langle \Phi_g | \left[ u_{00}^\dagger + u_{10} ^\dagger\cos(\alpha_g) + u_{01}^\dagger \sin(\alpha_g) \right] \left( U_{g+1}^{N_g-1} \right)^\dagger \hat{H} U_{g+1}^{N_g-1} \left[ u_{00} + u_{10} \cos(\alpha_g) + u_{01} \sin(\alpha_g) \right] | \Phi_g \rangle \\
&= E_{00} 
+  E_{10} \cos(\alpha_g) + E_{01} \sin(\alpha_g) 
+  E_{20} \cos(\alpha_g)^2 + E_{02} \sin(\alpha_g)^2
+  E_{11} \cos(\alpha_g) \sin(\alpha_g)
\;.
\end{split}
\end{equation}
Therefore, its gradient can be evaluated by the following shift rule:
\begin{equation}
\label{eq:partial_der}
\frac{\partial E}{\partial \alpha_g}(\alpha_g) = \sum_{\ell=1}^4 y_\ell E(\alpha_g + \Delta_\ell)
\;,\;
\vett{y} = \left( 1 , -1 , \frac{\sqrt{2}-2}{\sqrt{8}} , -\frac{\sqrt{2}-2}{\sqrt{8}} \right) 
\;,\;
\vett{\Delta} = \left( \frac{\pi}{4} , -\frac{\pi}{4} , \frac{\pi}{2} , -\frac{\pi}{2} \right) 
\;.
\end{equation}
Similarly, the second derivative reads $\frac{\partial^2 E}{\partial \alpha_g^2}(\alpha_g) = \sum_{\ell m=1}^4 y_\ell y_m E(\alpha_g + \Delta_\ell + \Delta_m)$.

If the energies $E(\alpha_g + \Delta_\ell)$ are measured with $S_\ell$ shots, due to Eq.~\eqref{eq:energy_variance}, the partial derivative in Eq.~\eqref{eq:partial_der} has variance
\begin{equation}
\mbox{Var}\left[ \frac{\partial E}{\partial \alpha_g}(\alpha_g) \right]
\leq \sum_{\ell=1}^4 \frac{y_\ell^2 \Lambda^2}{S_\ell} 
\to \frac{ \| \vett{y} \|_1^2 \Lambda^2}{ S }
\;,
\end{equation}
where $S = \sum_\ell S_\ell$ and, in the last step, we used the Lagrange multiplier technique. 

Measuring all the entries of the energy gradient with statistical uncertainty $\varepsilon$ thus requires:
\begin{displaymath}
\begin{array}{ll}
\mbox{circuits:} & 4 N_g (1+N_\gamma) = O(LN^3) \\
\\
\mbox{shots:}    & N_g \| \vett{y} \|_1^2 \Lambda^2 \varepsilon^{-2} = O(LN^2 \Lambda^2 \varepsilon^{-2})
\end{array}
\end{displaymath}

\subsection{Overlap matrix}

The overlap matrix in Eq.~\eqref{eq:overlap} can be written \cite{meyer2021fisher,wierichs2022general,stokes2020quantum} as the Hessian of the following function
\begin{equation}
\mathcal{S}_{gh} = \frac{1}{2} \frac{\partial^2 D}{\partial \delta_g \partial \delta_h}(\vett{\delta}=\vett{0})
\;,\;
D(\vett{\delta}) = 1 - \langle \Psi(\vett{\alpha}+\vett{\delta}) | \Psi(\vett{\alpha}) \rangle \langle \Psi(\vett{\alpha}) | \Psi(\vett{\alpha}+\vett{\delta}) \rangle
\;.
\end{equation}
Since $D(\vett{\delta})$ is the expectation value of the projector $|\Psi(\vett{\alpha}) \rangle \langle \Psi(\vett{\alpha})|$ over the state
$|\Psi(\vett{\alpha}+\vett{\delta}) \rangle$, it can be evaluated with a single quantum circuit of depth twice that of the LUCJ circuit, as the probability that a register of qubits prepared in $\hat{U}^\dagger(\vett{\alpha}) \hat{U}(\vett{\alpha}+\vett{\delta}) | \reference \rangle$ collapses onto $| \reference \rangle$ upon measurement,
\begin{equation}
D(\vett{\delta}) = | \langle \reference | \hat{U}^\dagger(\vett{\alpha}) \hat{U}(\vett{\alpha}+\vett{\delta}) | \reference \rangle |^2 \;.
\end{equation}
Furthermore, its second derivatives can be computed with the same shift rules described for the energy. The statistical uncertainty on $\mathcal{S}_{ij}$ is thus bounded by
\begin{equation}
\mbox{Var}[ \mathcal{S}_{gh} ]
= 
\sum_{\ell m} \frac{y_\ell^2 y_m^2}{S_{\ell m}}
\to 
\frac{ \| \vett{y} \|_1^4 }{S}
\;,
\end{equation}
where, in the equality, we recalled that $D$ is the expectation value of a projector (i.e. an operator with norm $1$) and, in the last step, we used the Lagrange multiplier technique. 

Measuring all the entries of the overlap matrix with statistical uncertainty $\varepsilon$ requires:
\begin{displaymath}
\begin{array}{ll}
\mbox{circuits:} & 16 N_g \frac{N_g+1}{2} = O(L^2 N^4) \\
\\
\mbox{shots:}    & N_g \frac{N_g+1}{2} \| \vett{y} \|_1^4 \varepsilon^{-2} = O(L^2 N^4 \varepsilon^{-2})
\end{array}
\end{displaymath}

However, recall that measuring $\mathcal{S}$ with shift rules requires executing a circuit of depth twice that of an LUCJ circuit, which may be undesirable on near-term devices. An alternative strategy is to measure a larger number of shallower circuits. To this end, let us observe that
\begin{enumerate}
\item $\hat{U}_g(\alpha_g) = e^{i \alpha_g B_g}$ for a Hermitian operator $B_g$, so that 
$\frac{\partial U_g}{\partial \alpha_g}(\alpha_g) = i B_g \hat{U}_g(\alpha_g)$.
\item $B_g = (1-Z)/2$ for $\mathsf{P}$ gates, $B_g = (1-Z_0)(1-Z_1)/4$ for $\mathsf{nn}$ gates, and $B_g = (XX+YY)/2$ for $\mathsf{XX+YY}$ gates. In all cases, $B_g = \sum_{\ell} b^g_\ell \sigma_\ell$ where $\| b^g \|_1 = 1$.
\item $B_g$ is either a 1-qubit operator, or a 2-qubit operator acting on neighboring qubits.
\item $| \Psi^g \rangle = \frac{\partial}{\partial {\alpha_g}} | \Psi \rangle = 
U_{g+1}^{N_g-1} i B_g U_0^g | \reference \rangle = U_{g+1}^{N_g-1} i B_g | \Phi_g \rangle$.
\end{enumerate}
As a result of these identities, one has that $\overline{S}_{gh} = \mbox{Re}[ \langle \Psi^g | \Psi^h \rangle - \langle \Psi^g | \Phi \rangle \langle \Psi | \Psi^h \rangle ]$ is a combination of
\begin{equation}
\begin{split}
\langle \Psi | \Psi^g \rangle &= i \langle \Phi_g | B_g | \Phi_g \rangle \;, \\
\langle \Psi^g | \Psi^g \rangle &= \phantom{i} \langle \Phi_g | B_g^2 | \Phi_g \rangle \;, \\
\langle \Psi^h | \Psi^g \rangle &= \mbox{Tr}\left[ B_h U_{g+1}^h B_g | \Phi_g \rangle \langle \Phi_g | \left( U_{g+1}^h \right)^\dagger \right] 
= \sum_{\ell m} b^g_\ell b^h_m \mbox{Tr}\left[ \sigma_m U_{g+1}^h \sigma_\ell | \Phi_g \rangle \langle \Phi_g | \left( U_{g+1}^h \right)^\dagger \right] \;,\; h>g \;.
\\
\end{split}
\end{equation}
The first two quantities are simple expectation values, whereas the third one is 
akin to a linear combination of correlation functions between Pauli operators. 

\begin{figure}
\includegraphics[width=0.8\textwidth]{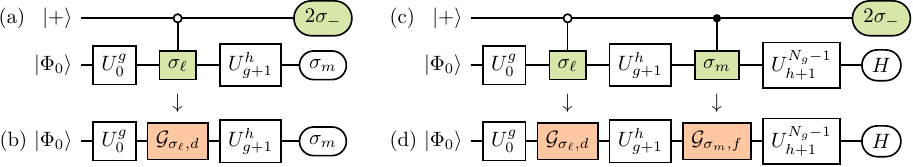}
\caption{(a) Quantum circuits to measure $\langle \Psi^h | \Psi^g \rangle$ using an ancilla and controlled-Pauli operations (with ancilla measurements, and controlled-Pauli marked in green and $\sigma_- = | 0 \rangle \langle 1 |$). (b) Quantum circuits to measure $\langle \Psi^h | \Psi^g \rangle$ using a set of quantum channels (with quantum channels marked orange).
(c,d) Equivalent circuits for $\langle \Psi^h | \hat{H} | \Psi^g \rangle$; the oval 
with the letter $H$ denotes the measurement of the Hamiltonian, which requires $(1+N_\gamma)$ circuits.}
\label{fig:circuits}
\end{figure}

The individual terms can be measured using the quantum circuits~\cite{somma2002simulating} shown in Figure~\ref{fig:circuits}a. These circuits require an ancilla qubit and a controlled-Pauli operation, which is undesirable on near-term devices. Ancillae and controlled-Pauli operation can be simulated ~\cite{mitarai2019methodology,mitarai2021overhead,mitarai2021constructing} by summing over a set of quantum channels (i.e. completely positive and trace-preserving maps between operator spaces) weighted by complex-valued coefficients.
In particular, as proved in Eq.~\eqref{eq:multiply} of the Appendix, the left-multiplication between a Pauli operator $\sigma$ and a density operator $\rho$ yields $\sigma \rho = \sum_{d=0}^3 L_d \, \mathcal{G}_{\sigma,d}[ \rho ]$ for suitable complex-valued coefficients $L_d$ and quantum channels $\mathcal{G}_{\sigma,d}$. As a result,
\begin{equation}
\label{eq:s_matrix}
\langle \Psi^h | \Psi^g \rangle
= 
\sum_{\ell m d} b^g_\ell b^h_m L_d \mbox{Tr}\left[ \sigma_m U_{g+1}^h \mathcal{G}_{\sigma_\ell,d}\Big[ | \Phi_g \rangle \langle \Phi_g | \Big] \left( U_{g+1}^h \right)^\dagger \right]
\end{equation}
which can be measured with the circuits in Figure~\ref{fig:circuits}b. The variance of this quantity is
\begin{equation}
\mbox{Var}\left( \langle \Psi^h | \Psi^g \rangle \right)
\leq
\sum_{\ell m d} \frac{ |b^g_\ell|^2 |b^h_m|^2 |L_d|^2 }{ S_{\ell m d} }
\to 
\frac{ \| b^g \|_1^2 \| b^h \|_1^2 \| L \|_1^2 }{ S } = \frac{9}{S}
\;.
\end{equation}
This technique based on quantum channels rather than shift rules  enables the computation of all elements of the overlap matrix with statistical uncertainty below $\varepsilon$ using fewer than
\begin{displaymath}
\begin{array}{ll}
\mbox{circuits:} & 4 N_g + 16 N_g + N_g \frac{N_g-1}{2} 64 = O(L^2 N^4) \\
\\
\mbox{shots:}    & \left( 2 N_g + N_g \frac{N_g-1}{2} 9 \right) \varepsilon^{-2} = O(L^2 N^4 \varepsilon^{-2} )
\end{array}
\end{displaymath}
We remark that the circuits have depth up to that of the original LUCJ circuit plus a modest overhead stemming from the 1- and 2-qubit $\mathcal{G}_{\sigma_\ell,d}$ channels.

\subsection{Hamiltonian matrix}

Unlike the overlap matrix, the Hamiltonian matrix in Eq.~\eqref{eq:hamiltonian}
is not the Hessian of an expectation value. Therefore, its evaluation using shift rules is less favorable due to the requirement of controlled operations, and we use the channel-based measurement technique of the previous subsection. One can write
\begin{equation}
\begin{split}
\langle \Psi^h | \hat{H} | \Psi^g \rangle 
&= 
\langle \reference | \left( U_0^h \right)^\dagger B_{h} \left( U_{h+1}^{N_g-1} \right)^\dagger \hat{H} U_{g+1}^{N_g-1} B_{g} U_0^g | \reference \rangle \\
&= \sum_{\ell m} b_\ell^g b_m^h 
\mbox{Tr}\left[ \hat{H} U_{h+1}^{N_g-1} \left( U_g^h \sigma_\ell | \Phi_g \rangle \langle \Phi_g | \left( U_g^h \right)^\dagger \sigma_m \right) \left( U_{h+1}^{N_g-1} \right)^\dagger \right]
\end{split}
\end{equation}
and measure this quantity with the circuits in Figure~\ref{fig:circuits}c, requiring ancillae and controlled-Pauli operations. These can be removed by sampling over quantum channels with a quasi-probability distribution, leading to the circuits in Figure~\ref{fig:circuits}d, and the following expression,
\begin{equation}
\begin{split}
\langle \Psi^h | \hat{H} | \Psi^g \rangle 
&= \sum_{\ell m d f } b_\ell^g b_m^h L_d R_f
\mbox{Tr}\left[ \hat{H} U_{h+1}^{N_g-1} 
\mathcal{G}_{\sigma_m,f}\Big[ 
U_g^h 
\mathcal{G}_{\sigma_\ell,d}\Big[ | \Phi_g \rangle \langle \Phi_g | \Big]
\left( U_g^h \right)^\dagger 
\Big] 
\left( U_{h+1}^{N_g-1} \right)^\dagger \right]
\;.
\end{split}
\end{equation}
The variance of this quantity is 
\begin{equation}
\begin{split}
\mbox{Var}( \langle \Psi^h | \hat{H} | \Psi^g \rangle )
\leq 
\sum_{\ell m d f } |b_\ell^g|^2 |b_m^h|^2 |L_d|^2 |R_f|^2
\frac{\Lambda^2}{S_{\ell m d f}}
\to
\frac{ \| b^g \|_1^2 \| b^h \|_1^2 \| L \|_1^2 \| R \|_1^2 \Lambda^2}{S}
=
\frac{81 \Lambda^2}{S}
\end{split}
\end{equation}
where in the right arrow  we have used the technique of Lagrange multipliers.
One can therefore compute all the elements of the Hamiltonian matrix with statistical uncertainty below $\varepsilon$ using fewer than
\begin{displaymath}
\begin{array}{ll}
\mbox{circuits:} & 256 (1+N_\gamma) N_g \frac{N_g+1}{2} = O(L^2 N^5) \\
\\
\mbox{shots:}    & N_g \frac{N_g+1}{2} \frac{81 \Lambda^2}{\varepsilon^2} = O(L^2 N^4 \Lambda^2 \varepsilon^{-2})
\end{array}
\end{displaymath}

\subsection{From circuit angles to LUCJ parameters}

So far, we focused on the measurement of the gradient vector and the overlap and Hamiltonian matrices as a function of the circuit angles $\vett{\alpha}$.  One may be interested in differentiating the energy with respect to the original parameters $\vett{\theta}$ (i.e. the elements of the $K$ and $J$ matrices).

Therefore, the gradient vector and the overlap and Hamiltonian matrices as a function of the circuit angles should be multiplied by the Jacobian and Hessian of a suitable transformation: for example, in order to differentiate with respect to the original parameters $\vett{\theta}$, one needs to consider the function $\vett{\alpha}(\vett{\theta})$, 
\begin{equation}
\frac{\partial E}{\partial \theta_i} = \sum_g \frac{\partial \alpha_g}{\partial \theta_i} \frac{\partial E}{\partial \alpha_g}
\;.
\end{equation}
The variance of the gradient vector with respect to the original parameters $\vett{\theta}$ is thus
\begin{equation}
\mbox{Var}\left( \frac{\partial E}{\partial \theta_i} \right)
= \sum_g \left| \frac{\partial \alpha_g}{\partial \theta_i} \right|^2 \mbox{Var}\left(\frac{\partial E}{\partial \alpha_g} \right)
\leq 
\sum_g \left| \frac{\partial \alpha_g}{\partial \theta_i} \right|^2 
\frac{ \| \vett{y} \|_1^2 \Lambda^2}{S}
=
\left\|  \frac{\partial \vett{\alpha}}{\partial \theta_i} \right\|_2^2 
\frac{ \| \vett{y} \|_1^2 \Lambda^2}{S}
\;.
\end{equation}
Therefore, to resolve all the components of $\frac{\partial E}{\partial \theta_i}$ within statistical uncertainty $\varepsilon$, one needs 
\begin{equation}
S = \max_i \left\| \frac{\partial \vett{\alpha}}{\partial \theta_i} \right\|_2^2 
\frac{ \| \vett{y} \|_1^2 \Lambda^2}{\varepsilon^2}
\end{equation}
shots, i.e., the number of shots needs to account for the gradient of the transformation $\vett{\alpha}(\vett{\theta})$. Similar formulas hold for the overlap and Hamiltonian matrices, and apply to both shift-rule-based and channel-based measurements.

While in principle one may differentiate the energy with respect to the circuit angles, rather than the LUCJ parameters, at least in the case of LUCJ these angles should not be treated as independent variables.
For example, for each $\mathsf{nn}$ gate implementing an $\alpha \alpha$ density-density interaction there is an $\mathsf{nn}$ gate with the same angle that implements a $\beta\beta$ density-density interaction. Similarly, for each $\mathsf{XX+YY}$ or $\mathsf{P}$ gate implementing a term of an $\alpha$ orbital rotation there is an $\mathsf{XX+YY}$ or $\mathsf{P}$ gate implementing a term of a $\beta$ orbital rotation.
To account for the relationship between different circuit angles, one can introduce a transformation $\vett{\alpha}( \vett{\tilde{\alpha}} )$ between all circuit angles and ``independent'' ones (arguably similar to the Z-matrix for molecular geometries): unlike $\vett{\alpha}(\vett{\theta})$, this is a linear and rather sparse function, potentially yielding a Jacobian with a smaller norm compared against $\vett{\alpha}(\vett{\theta})$.

\section{Discussion}

The LUCJ family of ansatzes is well-suited to describe both static and dynamic correlation.  The near-term limitation on the number of highly coherent qubits has required us to envision its possible use as an active space solver (i.e., focusing on the static correlation aspect of a larger problem).  We have multiple ideas for how to add active-inactive and inactive-inactive dynamic correlation – one particularly enticing research avenue involves tailored or externally-corrected coupled cluster,\cite{kinoshita2005coupled,morchen2020tailored,faulstich2019analysis,scheurer2024tailored} which would enable us to perform calculations in much larger basis sets.  An entirely different way to utilize the LUCJ ansatz in a hybrid quantum-classical computational protocol was recently proposed by some of us\cite{robledo2024chemistry}.  In short, the LUCJ probability distribution is used to generate raw configurations that, when corrected for number and other symmetries, can be used to generate subspaces for classical diagonalization of the Hamiltonian.

Numerical results from the above classical simulations with the LM optimizer show that it is a compelling alternative to L-BFGS-B for the optimization of the LUCJ ansatz, as the LM optimizer found lower LUCJ solutions at nearly all points across the dissociation curves of N$_2$ and C$_2$.  However, our  simulations did not include any potential sources of noise.   We note that while BFGS is one of the most widely used optimizers in classical simulations of quantum algorithms, it is not a suitable choice for performing, e.g., VQE on quantum hardware given its sensitivity to device noise: although the energy gradient is computed analytically the approximate Hessian update is obtained via finite difference~\cite{shi2022noise}.

\begin{figure}
\includegraphics[width=0.8\textwidth]{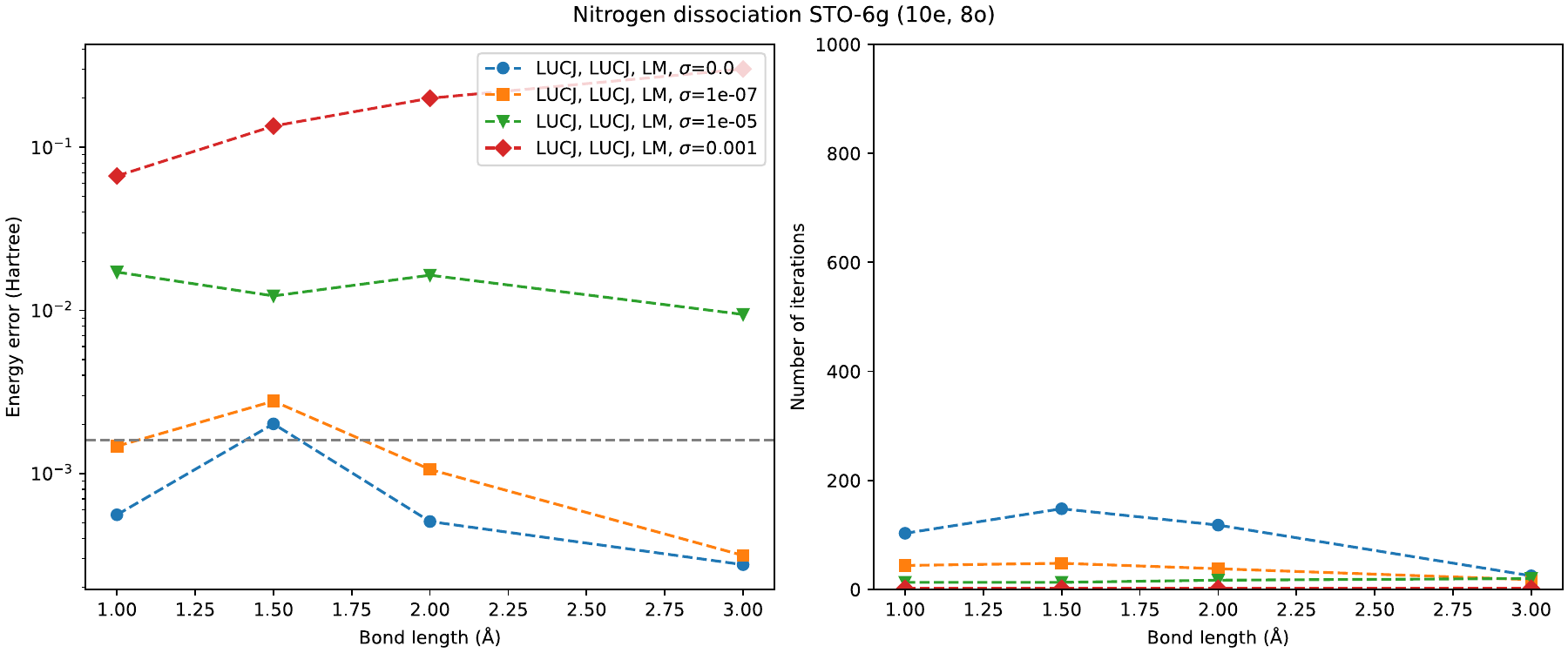}
\caption{Preliminary simulation of LM optimization of the LUCJ wavefunction of N$_2$ ($L=6$) with varying amounts of Gaussian noise added to the LM matrix elements.  $\sigma$ indicates the standard deviation of the Gaussian noise. The modified convergence criteria are described in the text.}
\label{fig:noise}
\end{figure}

Part of our motivation for focusing on the linear method (and stochastic reconfiguration) is that it has proved tremendously useful in the classical VMC community, where the LM matrix elements are also subject to statistical noise.  Indeed, in place of this statistical uncertainty on the LM elements due to MC sampling, our proposed quantum algorithm, when implemented on real hardware, will need to be robust against (at least) two types of quantum noise -- gate infidelity and shot noise.  Since we expect the effects of shot noise to be relatively more prominent in the context of LM-optimization of LUCJ-type wavefunctions, we have performed a preliminary study for a subset of the points in the N$_2$ dissociation curve, taking the number of layers, $L$, equal to 6.  To each element in the matrices in Equation 6, we have added Gaussian noise with standard deviation $\sigma$ = $10^{-x}$ where $x=3,5,7$, before solving the generalized eigenvalue problem.  As in the noiseless case, we variationally optimize the two hyper-parameters, $\alpha$ and $\xi$, at every LM iteration.  Note that these results have not employed a bootstrapping procedure, and thus cannot be directly compared to the results in Fig. 2 of the main text.  We have also changed the convergence criteria to be when either i) the gradient norm of the parameter update vector falls below max($10^{-5}$, $\sigma$), or ii) the total relative energy change falls below max($10^{-8}$, $\sigma$).  In Figure \ref{fig:noise}, we find that indeed, when the standard deviation of the Gaussian noise exceeds 10$^{-7}$ the efficacy of the LM optimizer deteriorates substantially.  But at 10$^{-7}$ and below, the optimized wavefunctions should be of comparable energy to the noiseless results.     
The number of shots implied by $\sigma < 10^{-7}$ is admittedly very large.  But we note that LM-optimization of the LUCJ is not the only algorithm that requires a large number of shots; indeed, essentially every other proposed near-term VQE-based algorithm does as well.  
With hardware capabilities improving at a rapid pace, we invite the reader to regard our work as a guide-post for where near-term quantum hardware needs to go, in order to feasibly support algorithms such as the one presented here. 

In a sense, the utility of the LM optimizer in VMC despite the requirement of a huge number of statistical samples for each matrix element makes us optimistic about its use in quantum variational optimization.  Just as in VMC, it is not infeasible to use quantum hardware resources in parallel to reduce the total wall-time required; indeed, one can now run multiple circuits on the same chip, because the size of a chip greatly exceeds the size of the circuits.  A second, complementary possibility is that reducing the dimension of the involved LM matrices by, e.g., selectively filtering out un-important parameters to be optimized\cite{garner2023improving} will permit the same accuracy resolution with a smaller number of shots.  Finally, we mention that techniques for reducing the required number of shots in quantum measurements are rapidly emerging\cite{arrasmith2020operator}, inspired by the concept of importance sampling.  
Taken together, the outlook for LM-optimization of LUCJ wavefunctions in the presence of shot noise is rosier than our preliminary numerical simulations with Gaussian noise would suggest.

\begin{center}
\begin{table}[th!]
\begin{tabular}{ |c|c|c| } 
\hline
$N_{\text{circuit}}$      & qSR & qLM  \\
\hline
\rule{0pt}{2.2ex}    Gradient & $\mathcal{O}(LN^3)$ & $\mathcal{O}(LN^3)$ \\ 
Overlap & $\mathcal{O}(L^2 N^4)$ & $\mathcal{O}(L^2 N^4)$  \\ 
Hamiltonian & N/A & $\mathcal{O}(L^2 N^5)$ \\[0.5ex] 
\hline
\end{tabular}
\caption{Asymptotic scaling of the number of circuits per iteration of qSR or qLM required to compute the gradient vector along with the overlap and Hamiltonian matrices.}
\label{tab:SummaryCircuits}
\end{table}
\end{center}

\begin{center}
\begin{table}[th!]
\begin{tabular}{ |c|c|c| } 
\hline
$N_{\text{shots}}$      & qSR & qLM  \\
\hline
\rule{0pt}{2.2ex}    Gradient & $\mathcal{O}(LN^2 \Lambda^2 \varepsilon^{-2})$ & $\mathcal{O}(LN^2 \Lambda^2 \varepsilon^{-2})$ \\ 
Overlap & $\mathcal{O}(L^2 N^4 \varepsilon^{-2})$ & $\mathcal{O}(L^2 N^4 \varepsilon^{-2})$  \\ 
Hamiltonian & N/A & $\mathcal{O}(L^2 N^4 \Lambda^2 \varepsilon^{-2})$ \\[0.5ex] 
\hline
\end{tabular}
\caption{Asymptotic scaling of the number of shots per iteration of qSR or qLM required to compute the gradient vector along with the overlap and Hamiltonian matrices.}
\label{tab:SummaryShots}
\end{table}
\end{center}

According to our analysis of the qSR and qLM algorithms, summarized in Tables \ref{tab:SummaryCircuits} and \ref{tab:SummaryShots}, qLM is more expensive as the number of measurements per iteration scales as a higher power of system size.
This is to be expected given that qSR requires only measurements of the overlap matrix elements, while the qLM in addition requires the Hamiltonian matrix elements.  
While we are not aware of a direct comparison of SR and the LM in the context of VMC, we expect that the qLM is likely to require fewer optimization iterations in the absence of measurement noise.  An additional aspect that ought to be investigated in future work is the dependence of the number of optimization iterations on the number of circuits/shots.

We note that numerical issues can, in principle, arise in both SR and the LM due to an ill-conditioned overlap matrix.  The analysis by Epperly, Lin, and Nakatsukasa\cite{epperly2022theory} which uses matrix perturbation theory to characterize the effects of noise on the quantum subspace expansion algorithm also applies to our Equations \ref{eq:lin_eq_1} and \ref{eq:hamiltonian}.  As formulated in, e.g., Theorem 2.7, there exist thresholding and subsequent truncation procedures that can be invoked to address the possible presence of an ill-conditioned overlap matrix.  
We also note another complementary strategy to the above, proposed very recently by Neuscamman\cite{garner2023improving}, which involves a judicious filtering of variational parameters in stochastic optimization procedures.  This essentially uses a selected-CI-like procedure within VMC to increase the effectiveness of noisy parameter optimization, which is likely to yield overlap matrices of smaller dimension and with fewer linear dependencies.

\subsection{Approximate qLM and qSR}

For both methods, the dominant computational cost originates from the orbital rotations in the LUCJ circuits.  Indeed, the number of variational parameters involved in the orbital rotation operators scales as $\mathcal{O}(N^2)$ as opposed to $\mathcal{O}(N)$ for the density-density interactions.
One potentially economical possibility to approximate qLM or qSR is to consider a restricted class of orbital rotations, e.g., those mapping to constant-depth Bogolyubov circuits. For such orbital rotations, the number of parameters scales as $\mathcal{O}(N)$, implying a substantial reduction of the number of parameters (and quantum gates).
Considering a restricted class of orbital rotations comes with an additional approximation to LUCJ, which is bound to affect its accuracy. The performance of this flavor of LUCJ will be considered in future studies.

Finally, the measurement analysis in the previous section assumes that the matrices $S$ and $H$ are dense.
An alternative to the application of eigenvalue cutoffs and/or a restriction to orbital rotations with constant-depth circuits involves sparsity-based approximations to qLM and qSR.
Block-LM has been combined with accelerated descent\cite{otis2019complementary}, and it is known that block-LM methods are able to treat up to 25,000 variational parameters~\cite{zhao2017blocked}.  SR with Krylov subspace methods avoids the need to construct full H and S matrices~\cite{neuscamman2012optimizing}.
Sparsity can be quantified at the start of the simulation with relatively low-accuracy measurements, and then employed to measure a subset of the variational parameters, thereby reducing the cost of qLM and qSR.


\subsection{On the importance of symmetries in wavefunction optimization}

Our numerical data suggests that the LM-optimized 
LUCJ ansatz can yield total energies within 1 kcal/mol of exact solutions across the dissociation curves of the strongly correlated diatomics N$_2$ and C$_2$.  However, even with our 3-fold bootstrapping scheme, we are unable to obtain perfectly smooth curves, as necessary for computing energy derivatives (vibrational spectra, response properties, etc.) and running dynamics simulations. The encountered discontinuities are likely due to the near-degeneracy of eigenstates of different symmetry. 

The LUCJ ansatz that we employ commutes with the $\hat{N}$ and $\hat{S}_z$ operators, though not with the $\hat{S}^2$ operator and any molecular point-group symmetries (e.g. $\mathrm{D}_{\infty h}$). As a result, it can (and in practice does) break spin and spatial symmetries, thus not yielding a wavefunction with consistent character along dissociation.
Several techniques were proposed to enforce symmetry on variationally
optimized wavefunctions, including the variation-after-projection 
approach~\cite{scuseria2011projected,jimenez2012projected,shi2014symmetry,qiu2017projected,qiu2018projected,papastathopoulos2023symmetry} wherein a symmetry-broken wavefunction $\Psi$
is projected onto an eigenspace of the symmetry operators with desired
eigenvalues through the application of an operator $\hat{\Pi}$, and the projected wavefunction $\hat{\Pi} | \Psi \rangle$ is variationally optimized. 
The implementation of variation-after-projection approaches on a quantum
computer is subtle and device-dependent, and depends on the nature of the symmetry group
through the mathematical structure of the projector $\hat{\Pi}$. 
Approximate projection schemes have also been proposed recently\cite{yen2019exact}.

A particularly important role in quantum computing is played by binary symmetries, i.e. symmetry groups generated by Pauli operators $P_1 \dots P_g$. The irreps of binary symmetries are labeled by binary strings 
$s \in \{\pm 1 \}^g$, and the projectors on such irreps are given by
\begin{equation}
\label{eq:z2_proj}
\hat{\Pi}_s = \prod_{k=1}^g \frac{I + s_k P_k}{2}
\;,\;
\end{equation}
i.e. projectors on the joint eigenspaces of $P_1 \dots P_g$ with eigenvalues $s_1 \dots s_g$. Important examples are the parities 
of the spin-up and spin-down particle numbers, 
in the standard Jordan-Wigner representation as
\begin{equation}
P_\sigma = \prod_{p=0}^{N-1} Z_{p\sigma} \;,\; \sigma \in \{\uparrow,\downarrow\}
\;,\;
\end{equation}
and the generators of many molecular point-group symmetries, notably $\mathrm{C_i}, \mathrm{C_s}, \mathrm{C_2}, \mathrm{C_{2h}}, \mathrm{C_{2v}}, \mathrm{D_2}, \mathrm{D_{2h}}$. For simplicity, here we illustrate the case of $\mathrm{C_i}$, whose generator reads in the Jordan-Wigner representation
\begin{equation}
P_{\mathrm{C_i}} = \prod_{p \in \mathrm{A_u}} Z_{p\uparrow} Z_{p\downarrow}
\;.
\end{equation}
In the previous equation, basis orbitals are assumed to be in either the $\mathrm{A_g}$ or $\mathrm{A_u}$ irrep, labeled by $+1$/$\-1$.

Projectors of the form Eq.~\eqref{eq:z2_proj} can be treated in different ways, for example: 
\begin{itemize}
    \item Symmetry generators can be mapped onto single-qubit Pauli operators using a Clifford transformation and removed from the computation via the so-called qubit tapering technique ~\cite{bravyi2017tapering}. This approach enforces binary symmetries exactly, and reduces the number of qubits by $g$. However, it requires modifying both unitary transformations (e.g. the LUCJ ansatz) and operators to measure (e.g. the Hamiltonian), in a way that may lead to increased gate count and circuit depth.
    \item Symmetry generators can be measured as the computation unfolds, with post-selection over measurement outcomes or feed-forward control to rotate the post-measurement state in the target symmetry sector, in what is essentially a form of quantum error correction ~\cite{gottesman2010introduction,shor1995scheme,aharonov1997fault,knill1998resilient,kitaev2003fault}. This approach does not require changing the structure of the quantum circuit of interest beyond the addition of mid-circuit measurements and feed-forward operations, but only allows correction of errors if these additional operations are sufficiently accurate ~\cite{gottesman2010introduction}.
    \item Projectors onto irreps of a binary symmetry group can be applied through final measurements, i.e. one can variationally optimize the cost function 
\begin{equation}
E(\theta) = \frac{\langle \Psi(\theta) | \hat{\Pi} \hat{H} \hat{\Pi} | \Psi(\theta) \rangle}{\langle \Psi(\theta) | \hat{\Pi} | \Psi(\theta) \rangle}
\;.
\end{equation}
The projector $\hat{\Pi} = 2^{-g} \sum_{\ell=0}^{g} \sum_{i_1< \dots <i_l} s_{i_1} \dots s_{i_l} P_{i_1} \dots P_{i_l}$ is easily written as a linear combination of $2^g$ Pauli operators, and thus can be measured along with the product $\hat{\Pi} \hat{H} \hat{\Pi} = \hat{H} \hat{\Pi}$ (where we have used the fact that $\hat{\Pi}$ commutes with the Hamiltonian and that it is idempotent),
upon writing also the Hamiltonian as a linear combination of Pauli operators.
The overhead of this approach is $2^g$ (imperfect) quantum measurements, but for most symmetry groups $g=O(1)$.  To give a few concrete examples of relevant $g$ values for binary symmetries:  for the parity symmetry, $g$ = 2; for $C_{2v}$, $g$ = 2; for $D_{2h}$, $g$ = 3; for binary translational symmetries in a 2x2x2 unit cell, $g$ = 3.
\end{itemize}  

\subsection{Constrained parameter optimization} 
Given the computational cost
of applying projectors on quantum devices, a reasonable approximation to
symmetry projection strategies is to constrain the parameter update to mitigate symmetry-breaking updates. Consider the case
of an observable $B(\vett{\theta}) = \langle \Psi(\vett{\theta}) | \hat{B} | \Psi(\vett{\theta}) \rangle$. Then,
\begin{equation}
B(\vett{\theta} + \Delta \vett{\theta}) = 
B(\vett{\theta}) 
+ 
\Delta \vett{\theta} \cdot \nabla B(\vett{\theta}) + O(\Delta \vett{\theta}^2)
\;.
\end{equation}
To conserve the symmetry $B$ to first order in $\Delta \vett{\theta}$,
one must require that $\Delta \vett{\theta}$ be orthogonal to $\nabla B(\vett{\theta})$.
Therefore, the solution of the linear equations Eq.~\eqref{eq:lin_eq_1} and Eq.~\eqref{eq:lin_eq_2} can be restricted in the subspace orthogonal to $\nabla B(\vett{\theta})$: one can expand $\Delta \vett{\theta} = \sum_i \Delta \theta_i {\bf{e}}_i$ on an orthonormal basis ${\bf{e}}_i$ whose first element is ${\bf{e}}_1 = \nabla B(\vett{\theta}) / \| \nabla B(\vett{\theta}) \|$ and require that $\Delta \theta_1 = 0$.
It should be noted that the proposed parameter change only conserves $B$ to first order in $\Delta \vett{\theta}$ and, in practice, symmetry-broken wavefunctions may still be produced as the variational optimization unfolds. To overcome this limitation, the update step can be rescaled to ensure the dominance of the first-order term, similarly to the rescaling parameter in Eq.~\eqref{eq:rescaling_1}.
This symmetry-constrained optimization procedure is suitable for the LM and SR, as the wavefunction is also expanded to first order in the variational parameters.

\section{Conclusions}

Since the advent of VQE, a primary focus in the field of using near-term hardware  for quantum chemistry has been ansatz design, i.e., the development of previously-unexplored wavefunction forms that attain exact or near exact correlation energies.  The LUCJ ansatz, possibly among others, is optimal in that it is i) hardware-friendly (free of SWAP gates), ii) systematically improvable, and iii) physically motivated (related to Hubbard physics and more general than UCCSD).  An important finding of this work is that the ansatz is but one part of the quantum solution.  An equally important part is how the wavefunction is optimized.  

Indeed, from a theorist's point of view, quantum hardware is compelling relative to classical approaches because it has the potential of realizing approximate wavefunction forms that are both variational (non-perturbative)  and size-extensive (in addition to other properties such as orbital invariance).  In this work we have shown through classical simulation that for two molecular examples of strong correlation, i.e. the dissociation of N$_2$ and C$_2$ diatomics, the same LUCJ ansatz permits multiple energy minima depending on the optimizer employed.  Specifically, even when using a relatively laborious bootstrapping procedure, the L-BFGS-B optimizer routinely converges on solutions that are higher in energy than those from the LM optimizer.  When running VQE on quantum hardware, the L-BFGS-B optimizer is unsuitable in practice, but we expect that the presence of noise would exacerbate the problem of finding the global energy minimum, especially in the regime of strong correlation where near-degeneracies abound.

Given the history of the classical VMC community, which has gravitated toward   optimizers such as the LM and SR due to their relative robustness to (in this case, stochastic) noise, we have investigated numerically the potential accuracy of LM-optimized LUCJ ansatzes via noiseless classical simulations, and find that in almost every case a lower energy LUCJ solution is found vs L-BFGS-B.  With $L$=6 and 10 for N$_2$ and C$_2$, respectively, the LUCJ energies deviate by less than 1 kcal/mol (1.6 milliHartrees) from the exact values at every point across the dissociation curves. Preliminary simulations including Gaussian shot noise provide some indication of the noise-tolerance of the LM optimizer.

We have derived and analyzed quantum algorithms for the LM and SR, which we refer to as qLM and qSR.  The key components include the calculation of the energy gradient vector with respect to the variational parameters, and the overlap and Hamiltonian matrices with respect to the wavefunction at a given iteration and its parameter derivatives.  Regarding the latter two matrices, the algorithms that we propose herein are found to be more economical than alternatives based on ``shift rules''.  We note that further analysis ideally combined with numerical studies would be useful to more realistically estimate the costs involved if differentiation with respect to LUCJ parameters, $\theta$, rather than the circuit angles, $\alpha$, derived therefrom is desired (though on quantum hardware the circuit angles are arguably the more directly relevant set of variational parameters).

It should be emphasized that classical VMC on the LUCJ ansatz (in general, orbital-space wavefunctions involving non-linear correlators)  is infeasible, with compute cost growing exponentially with system size.  In contrast, our qLM and qSR algorithms require numbers of quantum circuits and shots that scale polynomially with system size.  This is a significant example of a scientifically-relevant task for which quantum algorithms can outperform classical ones. 

With suitably robust second-order optimizers such as qSR and qLM, the LUCJ ansatz is able to achieve very accurate correlation energies in challenging regimes of electron correlation.  However, we find that there is no guarantee that the energies as a function of nuclear coordinate will evolve smoothly, such that energy derivatives and, e.g., empirical interatomic potentials can be obtained.  With this in mind, we propose a number of approaches -- viable on quantum hardware and in strongly correlated regimes -- to target states with a specific set of symmetries preserved.  The first involves symmetry projection (in the style of variation-after-projection) of binary symmetries, and the second involves symmetry-constrained updates of the variational parameters in the context of qSR and qLM.  These ideas merit further exploration, alongside approximations that would reduce the cost of each qSR or qLM iteration; possibilities include orbital rotation schemes requiring a linear (in system size) number of parameters, and approximations predicated on the sparsity of the overlap and Hamiltonian matrix elements in qSR and qLM.

\section{Acknowledgements}
\noindent We thank E. Neuscamman and L. Otis for helpful discussions.
J. Shee acknowledges support from the Robert A. Welch
Foundation, Award No. A24-0273-001.

\section{Data availability}

The data for this work is available at \url{https://zenodo.org/doi/10.5281/zenodo.11075045}.

\section{Code availability}

The source code for this work is available at \url{https://github.com/kevinsung/lucj-ffsim}.

\appendix

\section{Calculation of overlaps and matrix elements}

\subsection{Left- and right-multiplication formulas}

Given a density operator $\rho$ and a Pauli operator $\sigma$, the operators
\begin{equation}
\sigma \rho = \frac{1}{2} \{ \sigma, \rho \} + \frac{1}{2} [\sigma,\rho]
\;,\;
\rho \sigma = \frac{1}{2} \{ \sigma, \rho \} - \frac{1}{2} [\sigma,\rho]
\end{equation}
can be written in terms of quantum channels applied to the state $\rho$. First, defining
\begin{equation}
\begin{array}{ll}
\mathcal{G}_{\sigma,0}[\rho] = \frac{1+\sigma}{2} \rho \frac{1+\sigma}{2} \;,\; & 
\mathcal{G}_{\sigma,1}[\rho] = e^{-i \frac{\pi}{4} \sigma} \rho e^{ i \frac{\pi}{4} \sigma} \;,\; \\
\mathcal{G}_{\sigma,2}[\rho] = \frac{1-\sigma}{2} \rho \frac{1-\sigma}{2} \;,\; & 
\mathcal{G}_{\sigma,3}[\rho] = e^{ i \frac{\pi}{4} \sigma} \rho e^{-i \frac{\pi}{4} \sigma} \;,\; \\
\end{array}
\end{equation}
one immediately has
\begin{equation}
\label{eq:multiply}
\begin{split}
\sigma \rho &= 
\mathcal{G}_{\sigma,0}[\rho] - \mathcal{G}_{\sigma,2}[\rho] + 
\frac{\mathcal{G}_{\sigma,1}[\rho]-\mathcal{G}_{\sigma,3}[\rho]}{2i} = \sum_{p=0}^3 L_p \, \mathcal{G}_{\sigma,d}[\rho]
\;,\; \\
\rho \sigma &= 
\mathcal{G}_{\sigma,0}[\rho] - \mathcal{G}_{\sigma,2}[\rho] - 
\frac{\mathcal{G}_{\sigma,1}[\rho]-\mathcal{G}_{\sigma,3}[\rho]}{2i} = \sum_{p=0}^3 R_p \, \mathcal{G}_{\sigma,d}[\rho]
\;.
\end{split}
\end{equation}
Second, the maps $\mathcal{G}_{\sigma,1}$ and $\mathcal{G}_{\sigma,3}$ are unitary transformations, and thus  quantum channels. The maps $\mathcal{G}_{\sigma,0}$ and $\mathcal{G}_{\sigma,2}$ are defined by the projectors on the eigenspaces of $\sigma$ with eigenvalues $\pm 1$, so they can be implemented by projectively measuring $\sigma$ and post-selecting based on measurement outcomes $\pm 1$.

\section{Additional data}

In this section, we present additional data associated with the numerics presented in the main text.

\begin{figure}[h!]
\includegraphics[width=\textwidth]{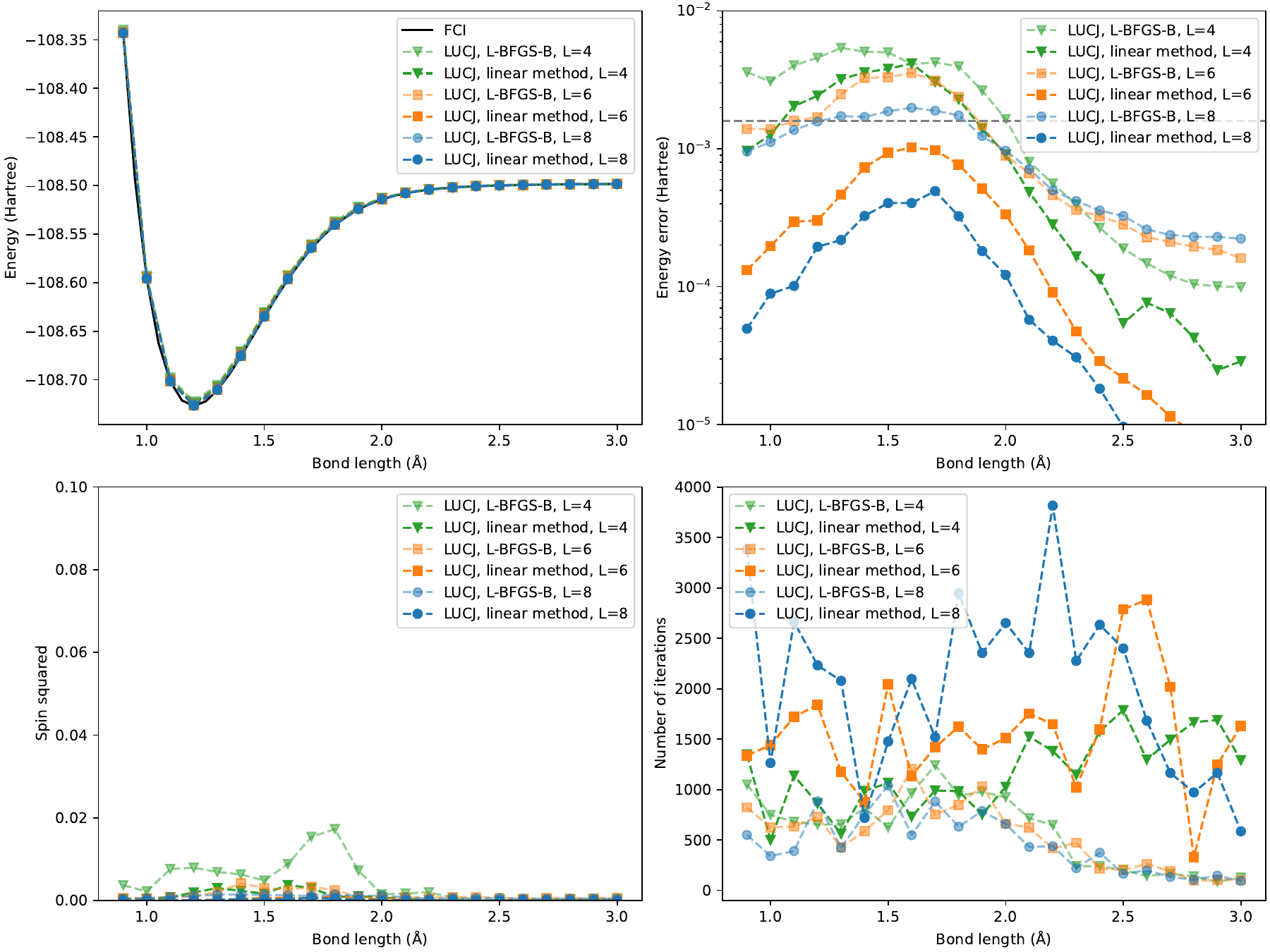}
\caption{Energy (upper left), error (upper right), spin squared (lower left), and number of optimization iterations (lower right) for  N$_2$, comparing L-BFGS-B and LM optimizers for LUCJ ansatzes with $L$=4 and 6. The number of iterations includes the total from all optimization runs from the bootstrapping procedure described in the main text.}
\label{fig:N2additional}
\end{figure}

\begin{figure}[h!]
\includegraphics[width=\textwidth]{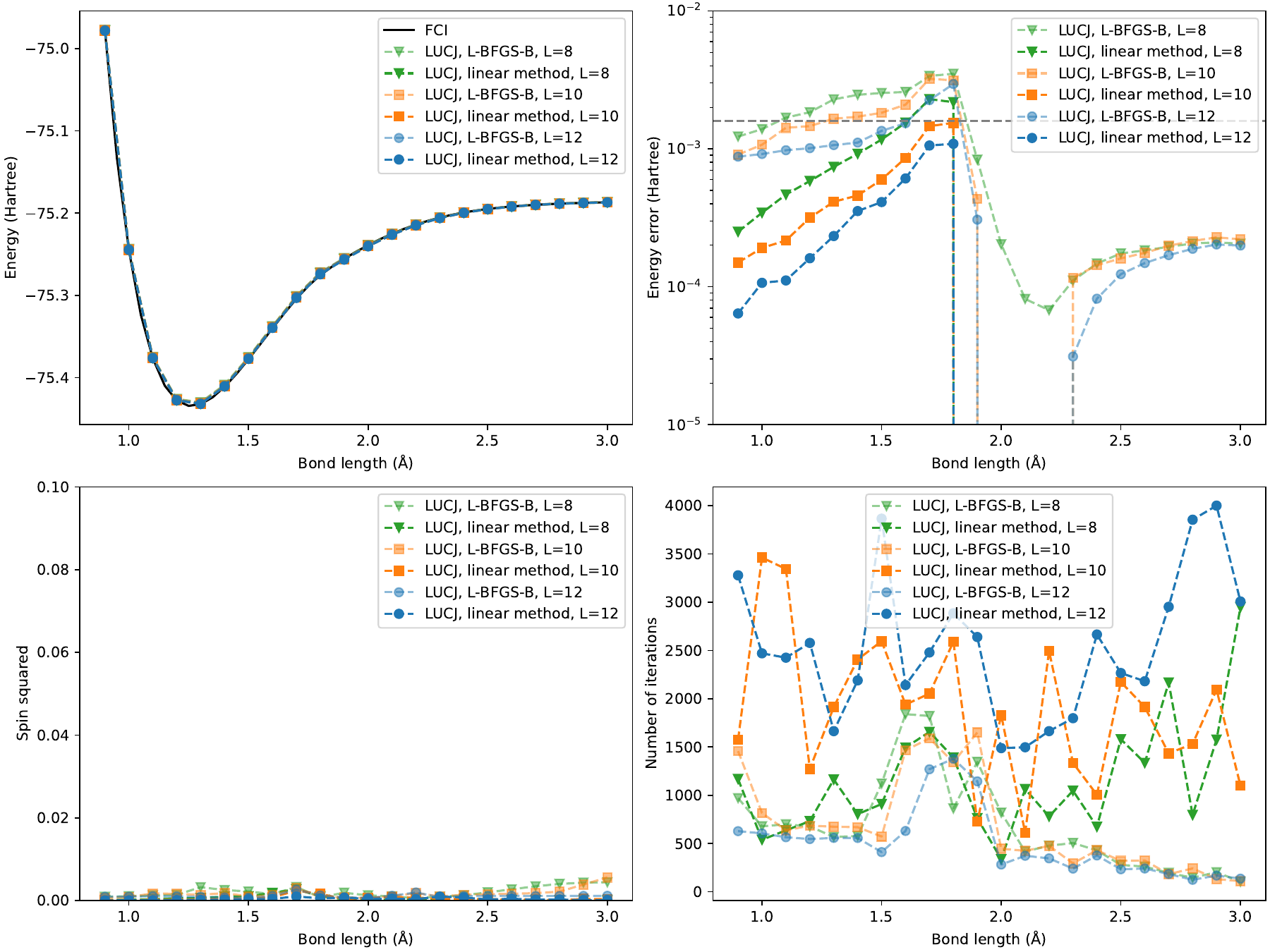}
\caption{Energy (upper left), error (upper right), spin squared (lower left), and number of optimization iterations (lower right) for  C$_2$, comparing L-BFGS-B and LM optimizers for LUCJ ansatzes with $L$=6 and 8. The number of iterations includes the total from all optimization runs from the bootstrapping procedure described in the main text.}
\label{fig:C2additional}
\end{figure}

Fig. \ref{fig:N2additional} shows the energy, error, spin squared, and number of optimization iterations for the dissociation curve of N$_2$. Fig. \ref{fig:C2additional} shows this data for C$_2$.

\begin{figure}[h!]
\includegraphics[width=\textwidth]{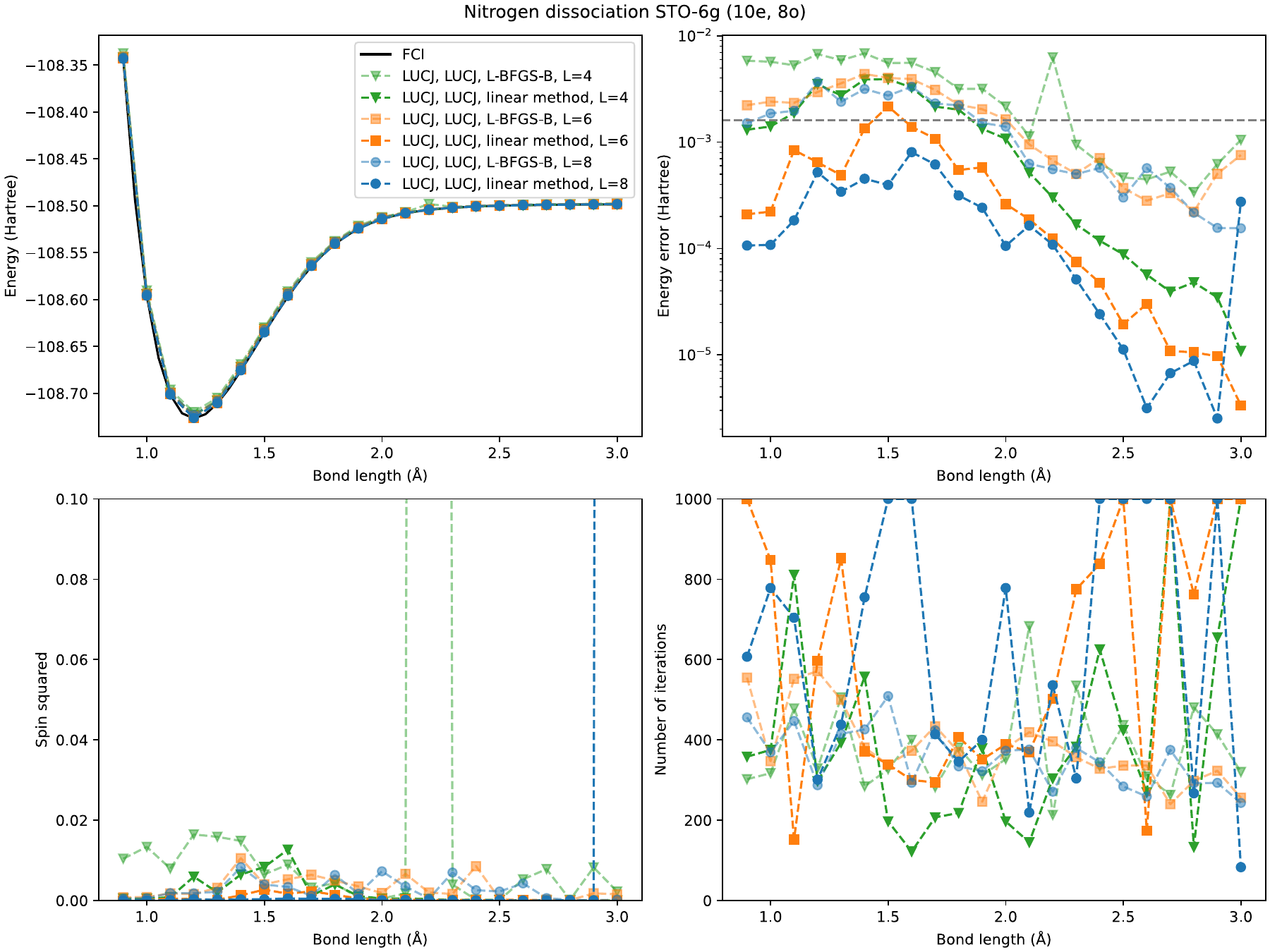}
\caption{Energy (upper left), error (upper right), spin squared (lower left), and number of optimization iterations (lower right) for  N$_2$, comparing L-BFGS-B and LM optimizers for LUCJ ansatzes with $L$=4 and 6. In this data, each data point was optimized independently from the initial parameters obtained from a truncated double-factorization of CCSD amplitudes, allowing a comparison between L-BFGS-B and LM in which the optimizers are always starting from the same initial parameters.}
\label{fig:N2nobootstrap}
\end{figure}

Fig. \ref{fig:N2nobootstrap} shows data for the dissociation curve of N$_2$ in which each data point was optimized independently from the initial parameters obtained from a truncated double-factorization of CCSD amplitudes, with no bootstrapping. The energies are worse than those obtained with bootstrapping, but this enables a comparison between L-BFGS-B and LM in which the optimizers are always starting from the same initial parameters.

\clearpage

\section*{References}
\bibliography{main}

\begin{thebibliography}{82}%
\makeatletter
\providecommand \@ifxundefined [1]{%
 \@ifx{#1\undefined}
}%
\providecommand \@ifnum [1]{%
 \ifnum #1\expandafter \@firstoftwo
 \else \expandafter \@secondoftwo
 \fi
}%
\providecommand \@ifx [1]{%
 \ifx #1\expandafter \@firstoftwo
 \else \expandafter \@secondoftwo
 \fi
}%
\providecommand \natexlab [1]{#1}%
\providecommand \enquote  [1]{``#1''}%
\providecommand \bibnamefont  [1]{#1}%
\providecommand \bibfnamefont [1]{#1}%
\providecommand \citenamefont [1]{#1}%
\providecommand \href@noop [0]{\@secondoftwo}%
\providecommand \href [0]{\begingroup \@sanitize@url \@href}%
\providecommand \@href[1]{\@@startlink{#1}\@@href}%
\providecommand \@@href[1]{\endgroup#1\@@endlink}%
\providecommand \@sanitize@url [0]{\catcode `\\12\catcode `\$12\catcode `\&12\catcode `\#12\catcode `\^12\catcode `\_12\catcode `\%12\relax}%
\providecommand \@@startlink[1]{}%
\providecommand \@@endlink[0]{}%
\providecommand \url  [0]{\begingroup\@sanitize@url \@url }%
\providecommand \@url [1]{\endgroup\@href {#1}{\urlprefix }}%
\providecommand \urlprefix  [0]{URL }%
\providecommand \Eprint [0]{\href }%
\providecommand \doibase [0]{https://doi.org/}%
\providecommand \selectlanguage [0]{\@gobble}%
\providecommand \bibinfo  [0]{\@secondoftwo}%
\providecommand \bibfield  [0]{\@secondoftwo}%
\providecommand \translation [1]{[#1]}%
\providecommand \BibitemOpen [0]{}%
\providecommand \bibitemStop [0]{}%
\providecommand \bibitemNoStop [0]{.\EOS\space}%
\providecommand \EOS [0]{\spacefactor3000\relax}%
\providecommand \BibitemShut  [1]{\csname bibitem#1\endcsname}%
\let\auto@bib@innerbib\@empty
\bibitem [{\citenamefont {Peruzzo}\ \emph {et~al.}(2014)\citenamefont {Peruzzo}, \citenamefont {McClean}, \citenamefont {Shadbolt}, \citenamefont {Yung}, \citenamefont {Zhou}, \citenamefont {Love}, \citenamefont {Aspuru-Guzik},\ and\ \citenamefont {O'Brien}}]{peruzzo2014variational}%
  \BibitemOpen
  \bibfield  {author} {\bibinfo {author} {\bibfnamefont {A.}~\bibnamefont {Peruzzo}}, \bibinfo {author} {\bibfnamefont {J.}~\bibnamefont {McClean}}, \bibinfo {author} {\bibfnamefont {P.}~\bibnamefont {Shadbolt}}, \bibinfo {author} {\bibfnamefont {M.-H.}\ \bibnamefont {Yung}}, \bibinfo {author} {\bibfnamefont {X.-Q.}\ \bibnamefont {Zhou}}, \bibinfo {author} {\bibfnamefont {P.~J.}\ \bibnamefont {Love}}, \bibinfo {author} {\bibfnamefont {A.}~\bibnamefont {Aspuru-Guzik}},\ and\ \bibinfo {author} {\bibfnamefont {J.~L.}\ \bibnamefont {O'Brien}},\ }\bibfield  {title} {\bibinfo {title} {A variational eigenvalue solver on a photonic quantum processor},\ }\href {https://doi.org/10.1038/ncomms5213} {\bibfield  {journal} {\bibinfo  {journal} {Nat. Commun}\ }\textbf {\bibinfo {volume} {5}},\ \bibinfo {pages} {4213} (\bibinfo {year} {2014})}\BibitemShut {NoStop}%
\bibitem [{\citenamefont {Anand}\ \emph {et~al.}(2022)\citenamefont {Anand}, \citenamefont {Schleich}, \citenamefont {Alperin-Lea}, \citenamefont {Jensen}, \citenamefont {Sim}, \citenamefont {D{\'\i}az-Tinoco}, \citenamefont {Kottmann}, \citenamefont {Degroote}, \citenamefont {Izmaylov},\ and\ \citenamefont {Aspuru-Guzik}}]{anand2022quantum}%
  \BibitemOpen
  \bibfield  {author} {\bibinfo {author} {\bibfnamefont {A.}~\bibnamefont {Anand}}, \bibinfo {author} {\bibfnamefont {P.}~\bibnamefont {Schleich}}, \bibinfo {author} {\bibfnamefont {S.}~\bibnamefont {Alperin-Lea}}, \bibinfo {author} {\bibfnamefont {P.~W.}\ \bibnamefont {Jensen}}, \bibinfo {author} {\bibfnamefont {S.}~\bibnamefont {Sim}}, \bibinfo {author} {\bibfnamefont {M.}~\bibnamefont {D{\'\i}az-Tinoco}}, \bibinfo {author} {\bibfnamefont {J.~S.}\ \bibnamefont {Kottmann}}, \bibinfo {author} {\bibfnamefont {M.}~\bibnamefont {Degroote}}, \bibinfo {author} {\bibfnamefont {A.~F.}\ \bibnamefont {Izmaylov}},\ and\ \bibinfo {author} {\bibfnamefont {A.}~\bibnamefont {Aspuru-Guzik}},\ }\bibfield  {title} {\bibinfo {title} {A quantum computing view on unitary coupled cluster theory},\ }\href {https://doi.org/10.1039/D1CS00932J} {\bibfield  {journal} {\bibinfo  {journal} {Chem. Soc. Rev}\ }\textbf {\bibinfo {volume} {51}},\ \bibinfo {pages} {1659} (\bibinfo {year} {2022})}\BibitemShut {NoStop}%
\bibitem [{\citenamefont {Matsuzawa}\ and\ \citenamefont {Kurashige}(2020)}]{matsuzawa2020jastrow}%
  \BibitemOpen
  \bibfield  {author} {\bibinfo {author} {\bibfnamefont {Y.}~\bibnamefont {Matsuzawa}}\ and\ \bibinfo {author} {\bibfnamefont {Y.}~\bibnamefont {Kurashige}},\ }\bibfield  {title} {\bibinfo {title} {Jastrow-type decomposition in quantum chemistry for low-depth quantum circuits},\ }\href {https://pubs.acs.org/doi/10.1021/acs.jctc.9b00963} {\bibfield  {journal} {\bibinfo  {journal} {J. Chem. Theory Comput}\ }\textbf {\bibinfo {volume} {16}},\ \bibinfo {pages} {944} (\bibinfo {year} {2020})}\BibitemShut {NoStop}%
\bibitem [{\citenamefont {Ryabinkin}\ \emph {et~al.}(2018)\citenamefont {Ryabinkin}, \citenamefont {Yen}, \citenamefont {Genin},\ and\ \citenamefont {Izmaylov}}]{ryabinkin2018qubit}%
  \BibitemOpen
  \bibfield  {author} {\bibinfo {author} {\bibfnamefont {I.~G.}\ \bibnamefont {Ryabinkin}}, \bibinfo {author} {\bibfnamefont {T.-C.}\ \bibnamefont {Yen}}, \bibinfo {author} {\bibfnamefont {S.~N.}\ \bibnamefont {Genin}},\ and\ \bibinfo {author} {\bibfnamefont {A.~F.}\ \bibnamefont {Izmaylov}},\ }\bibfield  {title} {\bibinfo {title} {Qubit coupled cluster method: a systematic approach to quantum chemistry on a quantum computer},\ }\href {https://pubs.acs.org/doi/10.1021/acs.jctc.8b00932} {\bibfield  {journal} {\bibinfo  {journal} {J. Chem. Theory Comput}\ }\textbf {\bibinfo {volume} {14}},\ \bibinfo {pages} {6317} (\bibinfo {year} {2018})}\BibitemShut {NoStop}%
\bibitem [{\citenamefont {Ryabinkin}\ \emph {et~al.}(2020)\citenamefont {Ryabinkin}, \citenamefont {Lang}, \citenamefont {Genin},\ and\ \citenamefont {Izmaylov}}]{ryabinkin2020iterative}%
  \BibitemOpen
  \bibfield  {author} {\bibinfo {author} {\bibfnamefont {I.~G.}\ \bibnamefont {Ryabinkin}}, \bibinfo {author} {\bibfnamefont {R.~A.}\ \bibnamefont {Lang}}, \bibinfo {author} {\bibfnamefont {S.~N.}\ \bibnamefont {Genin}},\ and\ \bibinfo {author} {\bibfnamefont {A.~F.}\ \bibnamefont {Izmaylov}},\ }\bibfield  {title} {\bibinfo {title} {Iterative qubit coupled cluster approach with efficient screening of generators},\ }\href@noop {} {\bibfield  {journal} {\bibinfo  {journal} {Journal of chemical theory and computation}\ }\textbf {\bibinfo {volume} {16}},\ \bibinfo {pages} {1055} (\bibinfo {year} {2020})}\BibitemShut {NoStop}%
\bibitem [{\citenamefont {Grimsley}\ \emph {et~al.}(2019)\citenamefont {Grimsley}, \citenamefont {Economou}, \citenamefont {Barnes},\ and\ \citenamefont {Mayhall}}]{grimsley2019adaptive}%
  \BibitemOpen
  \bibfield  {author} {\bibinfo {author} {\bibfnamefont {H.~R.}\ \bibnamefont {Grimsley}}, \bibinfo {author} {\bibfnamefont {S.~E.}\ \bibnamefont {Economou}}, \bibinfo {author} {\bibfnamefont {E.}~\bibnamefont {Barnes}},\ and\ \bibinfo {author} {\bibfnamefont {N.~J.}\ \bibnamefont {Mayhall}},\ }\bibfield  {title} {\bibinfo {title} {An adaptive variational algorithm for exact molecular simulations on a quantum computer},\ }\href@noop {} {\bibfield  {journal} {\bibinfo  {journal} {Nature communications}\ }\textbf {\bibinfo {volume} {10}},\ \bibinfo {pages} {3007} (\bibinfo {year} {2019})}\BibitemShut {NoStop}%
\bibitem [{\citenamefont {Tang}\ \emph {et~al.}(2021)\citenamefont {Tang}, \citenamefont {Shkolnikov}, \citenamefont {Barron}, \citenamefont {Grimsley}, \citenamefont {Mayhall}, \citenamefont {Barnes},\ and\ \citenamefont {Economou}}]{tang2021qubit}%
  \BibitemOpen
  \bibfield  {author} {\bibinfo {author} {\bibfnamefont {H.~L.}\ \bibnamefont {Tang}}, \bibinfo {author} {\bibfnamefont {V.}~\bibnamefont {Shkolnikov}}, \bibinfo {author} {\bibfnamefont {G.~S.}\ \bibnamefont {Barron}}, \bibinfo {author} {\bibfnamefont {H.~R.}\ \bibnamefont {Grimsley}}, \bibinfo {author} {\bibfnamefont {N.~J.}\ \bibnamefont {Mayhall}}, \bibinfo {author} {\bibfnamefont {E.}~\bibnamefont {Barnes}},\ and\ \bibinfo {author} {\bibfnamefont {S.~E.}\ \bibnamefont {Economou}},\ }\bibfield  {title} {\bibinfo {title} {qubit-adapt-vqe: An adaptive algorithm for constructing hardware-efficient ans{\"a}tze on a quantum processor},\ }\href@noop {} {\bibfield  {journal} {\bibinfo  {journal} {PRX Quantum}\ }\textbf {\bibinfo {volume} {2}},\ \bibinfo {pages} {020310} (\bibinfo {year} {2021})}\BibitemShut {NoStop}%
\bibitem [{\citenamefont {Motta}\ \emph {et~al.}(2023)\citenamefont {Motta}, \citenamefont {Sung}, \citenamefont {Whaley}, \citenamefont {Head-Gordon},\ and\ \citenamefont {Shee}}]{LUCJ}%
  \BibitemOpen
  \bibfield  {author} {\bibinfo {author} {\bibfnamefont {M.}~\bibnamefont {Motta}}, \bibinfo {author} {\bibfnamefont {K.~J.}\ \bibnamefont {Sung}}, \bibinfo {author} {\bibfnamefont {K.~B.}\ \bibnamefont {Whaley}}, \bibinfo {author} {\bibfnamefont {M.}~\bibnamefont {Head-Gordon}},\ and\ \bibinfo {author} {\bibfnamefont {J.}~\bibnamefont {Shee}},\ }\bibfield  {title} {\bibinfo {title} {Bridging physical intuition and hardware efficiency for correlated electronic states: the local unitary cluster jastrow ansatz for electronic structure},\ }\href {https://doi.org/10.1039/D3SC02516K} {\bibfield  {journal} {\bibinfo  {journal} {Chem. Sci}\ }\textbf {\bibinfo {volume} {14}},\ \bibinfo {pages} {11213} (\bibinfo {year} {2023})}\BibitemShut {NoStop}%
\bibitem [{\citenamefont {Ganoe}\ and\ \citenamefont {Shee}(2024)}]{ganoe2024notion}%
  \BibitemOpen
  \bibfield  {author} {\bibinfo {author} {\bibfnamefont {B.}~\bibnamefont {Ganoe}}\ and\ \bibinfo {author} {\bibfnamefont {J.}~\bibnamefont {Shee}},\ }\bibfield  {title} {\bibinfo {title} {On the notion of strong correlation in electronic structure theory},\ }\href {http://dx.doi.org/10.1039/D4FD00066H} {\bibfield  {journal} {\bibinfo  {journal} {Faraday Discuss}\ } (\bibinfo {year} {2024})}\BibitemShut {NoStop}%
\bibitem [{\citenamefont {Byrd}\ \emph {et~al.}(1995)\citenamefont {Byrd}, \citenamefont {Lu}, \citenamefont {Nocedal},\ and\ \citenamefont {Zhu}}]{byrd1995lbfgsb}%
  \BibitemOpen
  \bibfield  {author} {\bibinfo {author} {\bibfnamefont {R.~H.}\ \bibnamefont {Byrd}}, \bibinfo {author} {\bibfnamefont {P.}~\bibnamefont {Lu}}, \bibinfo {author} {\bibfnamefont {J.}~\bibnamefont {Nocedal}},\ and\ \bibinfo {author} {\bibfnamefont {C.}~\bibnamefont {Zhu}},\ }\bibfield  {title} {\bibinfo {title} {A limited memory algorithm for bound constrained optimization},\ }\href {https://doi.org/10.1137/0916069} {\bibfield  {journal} {\bibinfo  {journal} {SIAM J. Sci. Comp}\ }\textbf {\bibinfo {volume} {16}},\ \bibinfo {pages} {1190} (\bibinfo {year} {1995})}\BibitemShut {NoStop}%
\bibitem [{\citenamefont {Zhu}\ \emph {et~al.}(1997)\citenamefont {Zhu}, \citenamefont {Byrd}, \citenamefont {Lu},\ and\ \citenamefont {Nocedal}}]{zhu1997algorithm}%
  \BibitemOpen
  \bibfield  {author} {\bibinfo {author} {\bibfnamefont {C.}~\bibnamefont {Zhu}}, \bibinfo {author} {\bibfnamefont {R.~H.}\ \bibnamefont {Byrd}}, \bibinfo {author} {\bibfnamefont {P.}~\bibnamefont {Lu}},\ and\ \bibinfo {author} {\bibfnamefont {J.}~\bibnamefont {Nocedal}},\ }\bibfield  {title} {\bibinfo {title} {Algorithm 778: {L-BFGS-B}: {F}ortran subroutines for large-scale bound-constrained optimization},\ }\href {https://doi.org/10.1145/279232.279236} {\bibfield  {journal} {\bibinfo  {journal} {ACM Trans. Math. Softw.}\ }\textbf {\bibinfo {volume} {23}},\ \bibinfo {pages} {550–560} (\bibinfo {year} {1997})}\BibitemShut {NoStop}%
\bibitem [{\citenamefont {Chan}(2024)}]{chan2024quantum}%
  \BibitemOpen
  \bibfield  {author} {\bibinfo {author} {\bibfnamefont {G.~K.-L.}\ \bibnamefont {Chan}},\ }\bibfield  {title} {\bibinfo {title} {Quantum chemistry, classical heuristics, and quantum advantage},\ }\href@noop {} {\bibfield  {journal} {\bibinfo  {journal} {Faraday Discussions}\ } (\bibinfo {year} {2024})}\BibitemShut {NoStop}%
\bibitem [{\citenamefont {Foulkes}\ \emph {et~al.}(2001)\citenamefont {Foulkes}, \citenamefont {Mitas}, \citenamefont {Needs},\ and\ \citenamefont {Rajagopal}}]{foulkes2001quantum}%
  \BibitemOpen
  \bibfield  {author} {\bibinfo {author} {\bibfnamefont {W.}~\bibnamefont {Foulkes}}, \bibinfo {author} {\bibfnamefont {L.}~\bibnamefont {Mitas}}, \bibinfo {author} {\bibfnamefont {R.}~\bibnamefont {Needs}},\ and\ \bibinfo {author} {\bibfnamefont {G.}~\bibnamefont {Rajagopal}},\ }\bibfield  {title} {\bibinfo {title} {Quantum {{Monte Carlo}} simulations of solids},\ }\href {https://journals.aps.org/rmp/abstract/10.1103/RevModPhys.73.33} {\bibfield  {journal} {\bibinfo  {journal} {Rev. Mod. Phys}\ }\textbf {\bibinfo {volume} {73}},\ \bibinfo {pages} {33} (\bibinfo {year} {2001})}\BibitemShut {NoStop}%
\bibitem [{\citenamefont {Wei}\ and\ \citenamefont {Neuscamman}(2018)}]{wei2018reduced}%
  \BibitemOpen
  \bibfield  {author} {\bibinfo {author} {\bibfnamefont {H.}~\bibnamefont {Wei}}\ and\ \bibinfo {author} {\bibfnamefont {E.}~\bibnamefont {Neuscamman}},\ }\bibfield  {title} {\bibinfo {title} {Reduced scaling {Hilbert space variational {Monte Carlo}}},\ }\href {https://doi.org/10.1063/1.5047207} {\bibfield  {journal} {\bibinfo  {journal} {J. Chem. Phys}\ }\textbf {\bibinfo {volume} {149}},\ \bibinfo {pages} {184106} (\bibinfo {year} {2018})}\BibitemShut {NoStop}%
\bibitem [{\citenamefont {Sabzevari}\ and\ \citenamefont {Sharma}(2018)}]{sabzevari2018improved}%
  \BibitemOpen
  \bibfield  {author} {\bibinfo {author} {\bibfnamefont {I.}~\bibnamefont {Sabzevari}}\ and\ \bibinfo {author} {\bibfnamefont {S.}~\bibnamefont {Sharma}},\ }\bibfield  {title} {\bibinfo {title} {Improved speed and scaling in orbital space variational {{Monte Carlo}}},\ }\href {https://doi.org/10.1021/acs.jctc.8b00780} {\bibfield  {journal} {\bibinfo  {journal} {J. Chem. Theory Comput}\ }\textbf {\bibinfo {volume} {14}},\ \bibinfo {pages} {6276} (\bibinfo {year} {2018})}\BibitemShut {NoStop}%
\bibitem [{\citenamefont {Mahajan}\ and\ \citenamefont {Sharma}(2020)}]{mahajan2020efficient}%
  \BibitemOpen
  \bibfield  {author} {\bibinfo {author} {\bibfnamefont {A.}~\bibnamefont {Mahajan}}\ and\ \bibinfo {author} {\bibfnamefont {S.}~\bibnamefont {Sharma}},\ }\bibfield  {title} {\bibinfo {title} {{Efficient local energy evaluation for multi-Slater wave functions in orbital space quantum {Monte Carlo}}},\ }\href {https://doi.org/10.1063/5.0025055} {\bibfield  {journal} {\bibinfo  {journal} {J. Chem. Phys}\ }\textbf {\bibinfo {volume} {153}},\ \bibinfo {pages} {194108} (\bibinfo {year} {2020})}\BibitemShut {NoStop}%
\bibitem [{\citenamefont {Sorella}(2001)}]{SR_Sorella}%
  \BibitemOpen
  \bibfield  {author} {\bibinfo {author} {\bibfnamefont {S.}~\bibnamefont {Sorella}},\ }\bibfield  {title} {\bibinfo {title} {Generalized {Lanczos algorithm for variational quantum {Monte Carlo}}},\ }\href {https://doi.org/10.1103/PhysRevB.64.024512} {\bibfield  {journal} {\bibinfo  {journal} {Phys. Rev. B}\ }\textbf {\bibinfo {volume} {64}},\ \bibinfo {pages} {024512} (\bibinfo {year} {2001})}\BibitemShut {NoStop}%
\bibitem [{\citenamefont {Nightingale}\ and\ \citenamefont {Melik-Alaverdian}(2001)}]{nightingale2001optimization}%
  \BibitemOpen
  \bibfield  {author} {\bibinfo {author} {\bibfnamefont {M.}~\bibnamefont {Nightingale}}\ and\ \bibinfo {author} {\bibfnamefont {V.}~\bibnamefont {Melik-Alaverdian}},\ }\bibfield  {title} {\bibinfo {title} {Optimization of ground-and excited-state wave functions and {van der Waals} clusters},\ }\href {https://journals.aps.org/prl/abstract/10.1103/PhysRevLett.87.043401} {\bibfield  {journal} {\bibinfo  {journal} {Phys. Rev. Lett}\ }\textbf {\bibinfo {volume} {87}},\ \bibinfo {pages} {043401} (\bibinfo {year} {2001})}\BibitemShut {NoStop}%
\bibitem [{\citenamefont {Toulouse}\ and\ \citenamefont {Umrigar}(2007)}]{toulouse2007optimization}%
  \BibitemOpen
  \bibfield  {author} {\bibinfo {author} {\bibfnamefont {J.}~\bibnamefont {Toulouse}}\ and\ \bibinfo {author} {\bibfnamefont {C.~J.}\ \bibnamefont {Umrigar}},\ }\bibfield  {title} {\bibinfo {title} {{Optimization of quantum {Monte Carlo} wave functions by energy minimization}},\ }\href {https://doi.org/10.1063/1.2437215} {\bibfield  {journal} {\bibinfo  {journal} {J. Chem. Phys}\ }\textbf {\bibinfo {volume} {126}},\ \bibinfo {pages} {084102} (\bibinfo {year} {2007})}\BibitemShut {NoStop}%
\bibitem [{\citenamefont {Toulouse}\ and\ \citenamefont {Umrigar}(2008)}]{toulouse2008full}%
  \BibitemOpen
  \bibfield  {author} {\bibinfo {author} {\bibfnamefont {J.}~\bibnamefont {Toulouse}}\ and\ \bibinfo {author} {\bibfnamefont {C.}~\bibnamefont {Umrigar}},\ }\bibfield  {title} {\bibinfo {title} {Full optimization of {Jastrow--Slater} wave functions with application to the first-row atoms and homonuclear diatomic molecules},\ }\href {https://doi.org/10.1063/1.2908237} {\bibfield  {journal} {\bibinfo  {journal} {J. Chem. Phys}\ }\textbf {\bibinfo {volume} {128}},\ \bibinfo {pages} {174101} (\bibinfo {year} {2008})}\BibitemShut {NoStop}%
\bibitem [{\citenamefont {Umrigar}\ \emph {et~al.}(2007)\citenamefont {Umrigar}, \citenamefont {Toulouse}, \citenamefont {Filippi}, \citenamefont {Sorella},\ and\ \citenamefont {Hennig}}]{Umrigar_2007}%
  \BibitemOpen
  \bibfield  {author} {\bibinfo {author} {\bibfnamefont {C.~J.}\ \bibnamefont {Umrigar}}, \bibinfo {author} {\bibfnamefont {J.}~\bibnamefont {Toulouse}}, \bibinfo {author} {\bibfnamefont {C.}~\bibnamefont {Filippi}}, \bibinfo {author} {\bibfnamefont {S.}~\bibnamefont {Sorella}},\ and\ \bibinfo {author} {\bibfnamefont {R.~G.}\ \bibnamefont {Hennig}},\ }\bibfield  {title} {\bibinfo {title} {Alleviation of the fermion-sign problem by optimization of many-body wave functions},\ }\href {https://doi.org/10.1103/PhysRevLett.98.110201} {\bibfield  {journal} {\bibinfo  {journal} {Phys. Rev. Lett.}\ }\textbf {\bibinfo {volume} {98}},\ \bibinfo {pages} {110201} (\bibinfo {year} {2007})}\BibitemShut {NoStop}%
\bibitem [{\citenamefont {Umrigar}\ and\ \citenamefont {Filippi}(2005)}]{UmrigarFilippi2005}%
  \BibitemOpen
  \bibfield  {author} {\bibinfo {author} {\bibfnamefont {C.~J.}\ \bibnamefont {Umrigar}}\ and\ \bibinfo {author} {\bibfnamefont {C.}~\bibnamefont {Filippi}},\ }\bibfield  {title} {\bibinfo {title} {Energy and variance optimization of many-body wave functions},\ }\href {https://doi.org/10.1103/PhysRevLett.94.150201} {\bibfield  {journal} {\bibinfo  {journal} {Phys. Rev. Lett.}\ }\textbf {\bibinfo {volume} {94}},\ \bibinfo {pages} {150201} (\bibinfo {year} {2005})}\BibitemShut {NoStop}%
\bibitem [{\citenamefont {Motta}\ \emph {et~al.}(2021)\citenamefont {Motta}, \citenamefont {Ye}, \citenamefont {McClean}, \citenamefont {Li}, \citenamefont {Minnich}, \citenamefont {Babbush},\ and\ \citenamefont {Chan}}]{motta2021low}%
  \BibitemOpen
  \bibfield  {author} {\bibinfo {author} {\bibfnamefont {M.}~\bibnamefont {Motta}}, \bibinfo {author} {\bibfnamefont {E.}~\bibnamefont {Ye}}, \bibinfo {author} {\bibfnamefont {J.~R.}\ \bibnamefont {McClean}}, \bibinfo {author} {\bibfnamefont {Z.}~\bibnamefont {Li}}, \bibinfo {author} {\bibfnamefont {A.~J.}\ \bibnamefont {Minnich}}, \bibinfo {author} {\bibfnamefont {R.}~\bibnamefont {Babbush}},\ and\ \bibinfo {author} {\bibfnamefont {G.~K.}\ \bibnamefont {Chan}},\ }\bibfield  {title} {\bibinfo {title} {Low rank representations for quantum simulation of electronic structure},\ }\href {https://www.nature.com/articles/s41534-021-00416-z} {\bibfield  {journal} {\bibinfo  {journal} {npj Quantum Inf}\ }\textbf {\bibinfo {volume} {7}},\ \bibinfo {pages} {1} (\bibinfo {year} {2021})}\BibitemShut {NoStop}%
\bibitem [{\citenamefont {Kandala}\ \emph {et~al.}(2019)\citenamefont {Kandala}, \citenamefont {Temme}, \citenamefont {C{\'o}rcoles}, \citenamefont {Mezzacapo}, \citenamefont {Chow},\ and\ \citenamefont {Gambetta}}]{kandala2019error}%
  \BibitemOpen
  \bibfield  {author} {\bibinfo {author} {\bibfnamefont {A.}~\bibnamefont {Kandala}}, \bibinfo {author} {\bibfnamefont {K.}~\bibnamefont {Temme}}, \bibinfo {author} {\bibfnamefont {A.~D.}\ \bibnamefont {C{\'o}rcoles}}, \bibinfo {author} {\bibfnamefont {A.}~\bibnamefont {Mezzacapo}}, \bibinfo {author} {\bibfnamefont {J.~M.}\ \bibnamefont {Chow}},\ and\ \bibinfo {author} {\bibfnamefont {J.~M.}\ \bibnamefont {Gambetta}},\ }\bibfield  {title} {\bibinfo {title} {Error mitigation extends the computational reach of a noisy quantum processor},\ }\href@noop {} {\bibfield  {journal} {\bibinfo  {journal} {Nature}\ }\textbf {\bibinfo {volume} {567}},\ \bibinfo {pages} {491} (\bibinfo {year} {2019})}\BibitemShut {NoStop}%
\bibitem [{\citenamefont {Czarnik}\ \emph {et~al.}(2021)\citenamefont {Czarnik}, \citenamefont {Arrasmith}, \citenamefont {Coles},\ and\ \citenamefont {Cincio}}]{czarnik2021error}%
  \BibitemOpen
  \bibfield  {author} {\bibinfo {author} {\bibfnamefont {P.}~\bibnamefont {Czarnik}}, \bibinfo {author} {\bibfnamefont {A.}~\bibnamefont {Arrasmith}}, \bibinfo {author} {\bibfnamefont {P.~J.}\ \bibnamefont {Coles}},\ and\ \bibinfo {author} {\bibfnamefont {L.}~\bibnamefont {Cincio}},\ }\bibfield  {title} {\bibinfo {title} {Error mitigation with clifford quantum-circuit data},\ }\href {https://quantum-journal.org/papers/q-2021-11-26-592/} {\bibfield  {journal} {\bibinfo  {journal} {Quantum}\ }\textbf {\bibinfo {volume} {5}},\ \bibinfo {pages} {592} (\bibinfo {year} {2021})}\BibitemShut {NoStop}%
\bibitem [{\citenamefont {Nation}\ \emph {et~al.}(2021)\citenamefont {Nation}, \citenamefont {Kang}, \citenamefont {Sundaresan},\ and\ \citenamefont {Gambetta}}]{nation2021scalable}%
  \BibitemOpen
  \bibfield  {author} {\bibinfo {author} {\bibfnamefont {P.~D.}\ \bibnamefont {Nation}}, \bibinfo {author} {\bibfnamefont {H.}~\bibnamefont {Kang}}, \bibinfo {author} {\bibfnamefont {N.}~\bibnamefont {Sundaresan}},\ and\ \bibinfo {author} {\bibfnamefont {J.~M.}\ \bibnamefont {Gambetta}},\ }\bibfield  {title} {\bibinfo {title} {Scalable mitigation of measurement errors on quantum computers},\ }\href@noop {} {\bibfield  {journal} {\bibinfo  {journal} {PRX Quantum}\ }\textbf {\bibinfo {volume} {2}},\ \bibinfo {pages} {040326} (\bibinfo {year} {2021})}\BibitemShut {NoStop}%
\bibitem [{\citenamefont {Sun}\ \emph {et~al.}(2018)\citenamefont {Sun}, \citenamefont {Berkelbach}, \citenamefont {Blunt}, \citenamefont {Booth}, \citenamefont {Guo}, \citenamefont {Li}, \citenamefont {Liu}, \citenamefont {McClain}, \citenamefont {Sayfutyarova}, \citenamefont {Sharma} \emph {et~al.}}]{sun2018pyscf}%
  \BibitemOpen
  \bibfield  {author} {\bibinfo {author} {\bibfnamefont {Q.}~\bibnamefont {Sun}}, \bibinfo {author} {\bibfnamefont {T.~C.}\ \bibnamefont {Berkelbach}}, \bibinfo {author} {\bibfnamefont {N.~S.}\ \bibnamefont {Blunt}}, \bibinfo {author} {\bibfnamefont {G.~H.}\ \bibnamefont {Booth}}, \bibinfo {author} {\bibfnamefont {S.}~\bibnamefont {Guo}}, \bibinfo {author} {\bibfnamefont {Z.}~\bibnamefont {Li}}, \bibinfo {author} {\bibfnamefont {J.}~\bibnamefont {Liu}}, \bibinfo {author} {\bibfnamefont {J.~D.}\ \bibnamefont {McClain}}, \bibinfo {author} {\bibfnamefont {E.~R.}\ \bibnamefont {Sayfutyarova}}, \bibinfo {author} {\bibfnamefont {S.}~\bibnamefont {Sharma}}, \emph {et~al.},\ }\bibfield  {title} {\bibinfo {title} {{PySCF: the Python-based simulations of chemistry framework}},\ }\href {https://wires.onlinelibrary.wiley.com/doi/abs/10.1002/wcms.1340} {\bibfield  {journal} {\bibinfo  {journal} {WIREs Comput. Mol. Sci}\ }\textbf {\bibinfo {volume} {8}},\ \bibinfo {pages} {e1340} (\bibinfo {year} {2018})}\BibitemShut
  {NoStop}%
\bibitem [{\citenamefont {Sun}\ \emph {et~al.}(2020)\citenamefont {Sun}, \citenamefont {Zhang}, \citenamefont {Banerjee}, \citenamefont {Bao}, \citenamefont {Barbry}, \citenamefont {Blunt}, \citenamefont {Bogdanov}, \citenamefont {Booth}, \citenamefont {Chen}, \citenamefont {Cui} \emph {et~al.}}]{sun2020recent}%
  \BibitemOpen
  \bibfield  {author} {\bibinfo {author} {\bibfnamefont {Q.}~\bibnamefont {Sun}}, \bibinfo {author} {\bibfnamefont {X.}~\bibnamefont {Zhang}}, \bibinfo {author} {\bibfnamefont {S.}~\bibnamefont {Banerjee}}, \bibinfo {author} {\bibfnamefont {P.}~\bibnamefont {Bao}}, \bibinfo {author} {\bibfnamefont {M.}~\bibnamefont {Barbry}}, \bibinfo {author} {\bibfnamefont {N.~S.}\ \bibnamefont {Blunt}}, \bibinfo {author} {\bibfnamefont {N.~A.}\ \bibnamefont {Bogdanov}}, \bibinfo {author} {\bibfnamefont {G.~H.}\ \bibnamefont {Booth}}, \bibinfo {author} {\bibfnamefont {J.}~\bibnamefont {Chen}}, \bibinfo {author} {\bibfnamefont {Z.-H.}\ \bibnamefont {Cui}}, \emph {et~al.},\ }\bibfield  {title} {\bibinfo {title} {Recent developments in the {PySCF} program package},\ }\href {https://doi.org/10.1063/5.0006074} {\bibfield  {journal} {\bibinfo  {journal} {J. Chem. Phys}\ }\textbf {\bibinfo {volume} {153}},\ \bibinfo {pages} {024109} (\bibinfo {year} {2020})}\BibitemShut {NoStop}%
\bibitem [{\citenamefont {{The ffsim developers}}(2024)}]{ffsim}%
  \BibitemOpen
  \bibfield  {author} {\bibinfo {author} {\bibnamefont {{The ffsim developers}}},\ }\href@noop {} {\bibinfo {title} {{ffsim: Faster simulations of fermionic quantum circuits}}} (\bibinfo {year} {2024}),\ \bibinfo {note} {{\url{https://github.com/qiskit-community/ffsim}}}\BibitemShut {NoStop}%
\bibitem [{\citenamefont {{The SciPy 1.0 contributors}}(2020)}]{scipy}%
  \BibitemOpen
  \bibfield  {author} {\bibinfo {author} {\bibnamefont {{The SciPy 1.0 contributors}}},\ }\bibfield  {title} {\bibinfo {title} {{{SciPy} 1.0: Fundamental Algorithms for Scientific Computing in Python}},\ }\href {https://doi.org/10.1038/s41592-019-0686-2} {\bibfield  {journal} {\bibinfo  {journal} {Nat. Methods}\ }\textbf {\bibinfo {volume} {17}},\ \bibinfo {pages} {261} (\bibinfo {year} {2020})}\BibitemShut {NoStop}%
\bibitem [{\citenamefont {Fan}\ and\ \citenamefont {Piecuch}(2006)}]{fan2006usefulness}%
  \BibitemOpen
  \bibfield  {author} {\bibinfo {author} {\bibfnamefont {P.-D.}\ \bibnamefont {Fan}}\ and\ \bibinfo {author} {\bibfnamefont {P.}~\bibnamefont {Piecuch}},\ }\bibfield  {title} {\bibinfo {title} {The usefulness of exponential wave function expansions employing one-and two-body cluster operators in electronic structure theory: the extended and generalized coupled-cluster methods},\ }\href {https://www.sciencedirect.com/science/article/abs/pii/S0065327606510019} {\bibfield  {journal} {\bibinfo  {journal} {Adv. Quantum Chem}\ }\textbf {\bibinfo {volume} {51}},\ \bibinfo {pages} {1} (\bibinfo {year} {2006})}\BibitemShut {NoStop}%
\bibitem [{\citenamefont {Bulik}\ \emph {et~al.}(2015)\citenamefont {Bulik}, \citenamefont {Henderson},\ and\ \citenamefont {Scuseria}}]{bulik2015can}%
  \BibitemOpen
  \bibfield  {author} {\bibinfo {author} {\bibfnamefont {I.~W.}\ \bibnamefont {Bulik}}, \bibinfo {author} {\bibfnamefont {T.~M.}\ \bibnamefont {Henderson}},\ and\ \bibinfo {author} {\bibfnamefont {G.~E.}\ \bibnamefont {Scuseria}},\ }\bibfield  {title} {\bibinfo {title} {Can single-reference coupled cluster theory describe static correlation?},\ }\href {https://doi.org/10.1021/acs.jctc.5b00422} {\bibfield  {journal} {\bibinfo  {journal} {J. Chem. Theory Comput}\ }\textbf {\bibinfo {volume} {11}},\ \bibinfo {pages} {3171} (\bibinfo {year} {2015})}\BibitemShut {NoStop}%
\bibitem [{\citenamefont {Abrams}\ and\ \citenamefont {Sherrill}(2004)}]{abrams2004full}%
  \BibitemOpen
  \bibfield  {author} {\bibinfo {author} {\bibfnamefont {M.~L.}\ \bibnamefont {Abrams}}\ and\ \bibinfo {author} {\bibfnamefont {C.~D.}\ \bibnamefont {Sherrill}},\ }\bibfield  {title} {\bibinfo {title} {{Full configuration interaction potential energy curves for the $X^1 \Sigma^+_g$, $B^1 \Delta_g$, and $B^{\prime 1} \Sigma^+_g$ states of C$_2$: a challenge for approximate methods}},\ }\href {https://doi.org/10.1063/1.1804498} {\bibfield  {journal} {\bibinfo  {journal} {J. Chem. Phys}\ }\textbf {\bibinfo {volume} {121}},\ \bibinfo {pages} {9211} (\bibinfo {year} {2004})}\BibitemShut {NoStop}%
\bibitem [{\citenamefont {Purwanto}\ \emph {et~al.}(2009)\citenamefont {Purwanto}, \citenamefont {Zhang},\ and\ \citenamefont {Krakauer}}]{purwanto2009excited}%
  \BibitemOpen
  \bibfield  {author} {\bibinfo {author} {\bibfnamefont {W.}~\bibnamefont {Purwanto}}, \bibinfo {author} {\bibfnamefont {S.}~\bibnamefont {Zhang}},\ and\ \bibinfo {author} {\bibfnamefont {H.}~\bibnamefont {Krakauer}},\ }\bibfield  {title} {\bibinfo {title} {Excited state calculations using phaseless auxiliary-field quantum {{Monte Carlo}: Potential energy curves of low-lying C$_2$} singlet states},\ }\href {https://doi.org/10.1063/1.3077920} {\bibfield  {journal} {\bibinfo  {journal} {J. Chem. Phys}\ }\textbf {\bibinfo {volume} {130}},\ \bibinfo {pages} {094107} (\bibinfo {year} {2009})}\BibitemShut {NoStop}%
\bibitem [{\citenamefont {Booth}\ \emph {et~al.}(2011)\citenamefont {Booth}, \citenamefont {Cleland}, \citenamefont {Thom},\ and\ \citenamefont {Alavi}}]{booth2011breaking}%
  \BibitemOpen
  \bibfield  {author} {\bibinfo {author} {\bibfnamefont {G.~H.}\ \bibnamefont {Booth}}, \bibinfo {author} {\bibfnamefont {D.}~\bibnamefont {Cleland}}, \bibinfo {author} {\bibfnamefont {A.~J.}\ \bibnamefont {Thom}},\ and\ \bibinfo {author} {\bibfnamefont {A.}~\bibnamefont {Alavi}},\ }\bibfield  {title} {\bibinfo {title} {Breaking the carbon dimer: The challenges of multiple bond dissociation with full configuration interaction quantum {{Monte Carlo}} methods},\ }\href {https://doi.org/10.1063/1.3624383} {\bibfield  {journal} {\bibinfo  {journal} {J. Chem. Phys}\ }\textbf {\bibinfo {volume} {135}},\ \bibinfo {pages} {084104} (\bibinfo {year} {2011})}\BibitemShut {NoStop}%
\bibitem [{\citenamefont {Huggins}\ \emph {et~al.}(2021)\citenamefont {Huggins}, \citenamefont {McClean}, \citenamefont {Rubin}, \citenamefont {Jiang}, \citenamefont {Wiebe}, \citenamefont {Whaley},\ and\ \citenamefont {Babbush}}]{huggins2021efficient}%
  \BibitemOpen
  \bibfield  {author} {\bibinfo {author} {\bibfnamefont {W.~J.}\ \bibnamefont {Huggins}}, \bibinfo {author} {\bibfnamefont {J.~R.}\ \bibnamefont {McClean}}, \bibinfo {author} {\bibfnamefont {N.~C.}\ \bibnamefont {Rubin}}, \bibinfo {author} {\bibfnamefont {Z.}~\bibnamefont {Jiang}}, \bibinfo {author} {\bibfnamefont {N.}~\bibnamefont {Wiebe}}, \bibinfo {author} {\bibfnamefont {K.~B.}\ \bibnamefont {Whaley}},\ and\ \bibinfo {author} {\bibfnamefont {R.}~\bibnamefont {Babbush}},\ }\bibfield  {title} {\bibinfo {title} {Efficient and noise resilient measurements for quantum chemistry on near-term quantum computers},\ }\href {https://www.nature.com/articles/s41534-020-00341-7} {\bibfield  {journal} {\bibinfo  {journal} {npj Quantum Inf}\ }\textbf {\bibinfo {volume} {7}},\ \bibinfo {pages} {1} (\bibinfo {year} {2021})}\BibitemShut {NoStop}%
\bibitem [{\citenamefont {Choi}\ \emph {et~al.}(2023)\citenamefont {Choi}, \citenamefont {Loaiza},\ and\ \citenamefont {Izmaylov}}]{choi2023fluid}%
  \BibitemOpen
  \bibfield  {author} {\bibinfo {author} {\bibfnamefont {S.}~\bibnamefont {Choi}}, \bibinfo {author} {\bibfnamefont {I.}~\bibnamefont {Loaiza}},\ and\ \bibinfo {author} {\bibfnamefont {A.~F.}\ \bibnamefont {Izmaylov}},\ }\bibfield  {title} {\bibinfo {title} {Fluid fermionic fragments for optimizing quantum measurements of electronic hamiltonians in the variational quantum eigensolver},\ }\href@noop {} {\bibfield  {journal} {\bibinfo  {journal} {Quantum}\ }\textbf {\bibinfo {volume} {7}},\ \bibinfo {pages} {889} (\bibinfo {year} {2023})}\BibitemShut {NoStop}%
\bibitem [{\citenamefont {Choi}\ \emph {et~al.}(2022)\citenamefont {Choi}, \citenamefont {Yen},\ and\ \citenamefont {Izmaylov}}]{choi2022improving}%
  \BibitemOpen
  \bibfield  {author} {\bibinfo {author} {\bibfnamefont {S.}~\bibnamefont {Choi}}, \bibinfo {author} {\bibfnamefont {T.-C.}\ \bibnamefont {Yen}},\ and\ \bibinfo {author} {\bibfnamefont {A.~F.}\ \bibnamefont {Izmaylov}},\ }\bibfield  {title} {\bibinfo {title} {Improving quantum measurements by introducing “ghost” pauli products},\ }\href@noop {} {\bibfield  {journal} {\bibinfo  {journal} {Journal of Chemical Theory and Computation}\ }\textbf {\bibinfo {volume} {18}},\ \bibinfo {pages} {7394} (\bibinfo {year} {2022})}\BibitemShut {NoStop}%
\bibitem [{\citenamefont {Yen}\ \emph {et~al.}(2023)\citenamefont {Yen}, \citenamefont {Ganeshram},\ and\ \citenamefont {Izmaylov}}]{yen2023deterministic}%
  \BibitemOpen
  \bibfield  {author} {\bibinfo {author} {\bibfnamefont {T.-C.}\ \bibnamefont {Yen}}, \bibinfo {author} {\bibfnamefont {A.}~\bibnamefont {Ganeshram}},\ and\ \bibinfo {author} {\bibfnamefont {A.~F.}\ \bibnamefont {Izmaylov}},\ }\bibfield  {title} {\bibinfo {title} {Deterministic improvements of quantum measurements with grouping of compatible operators, non-local transformations, and covariance estimates},\ }\href@noop {} {\bibfield  {journal} {\bibinfo  {journal} {npj Quantum Information}\ }\textbf {\bibinfo {volume} {9}},\ \bibinfo {pages} {14} (\bibinfo {year} {2023})}\BibitemShut {NoStop}%
\bibitem [{\citenamefont {Jena}\ \emph {et~al.}(2022)\citenamefont {Jena}, \citenamefont {Genin},\ and\ \citenamefont {Mosca}}]{jena2022optimization}%
  \BibitemOpen
  \bibfield  {author} {\bibinfo {author} {\bibfnamefont {A.}~\bibnamefont {Jena}}, \bibinfo {author} {\bibfnamefont {S.~N.}\ \bibnamefont {Genin}},\ and\ \bibinfo {author} {\bibfnamefont {M.}~\bibnamefont {Mosca}},\ }\bibfield  {title} {\bibinfo {title} {Optimization of variational-quantum-eigensolver measurement by partitioning pauli operators using multiqubit clifford gates on noisy intermediate-scale quantum hardware},\ }\href@noop {} {\bibfield  {journal} {\bibinfo  {journal} {Physical Review A}\ }\textbf {\bibinfo {volume} {106}},\ \bibinfo {pages} {042443} (\bibinfo {year} {2022})}\BibitemShut {NoStop}%
\bibitem [{\citenamefont {Huang}\ \emph {et~al.}(2021)\citenamefont {Huang}, \citenamefont {Kueng},\ and\ \citenamefont {Preskill}}]{huang2021efficient}%
  \BibitemOpen
  \bibfield  {author} {\bibinfo {author} {\bibfnamefont {H.-Y.}\ \bibnamefont {Huang}}, \bibinfo {author} {\bibfnamefont {R.}~\bibnamefont {Kueng}},\ and\ \bibinfo {author} {\bibfnamefont {J.}~\bibnamefont {Preskill}},\ }\bibfield  {title} {\bibinfo {title} {Efficient estimation of pauli observables by derandomization},\ }\href@noop {} {\bibfield  {journal} {\bibinfo  {journal} {Physical review letters}\ }\textbf {\bibinfo {volume} {127}},\ \bibinfo {pages} {030503} (\bibinfo {year} {2021})}\BibitemShut {NoStop}%
\bibitem [{\citenamefont {Beebe}\ and\ \citenamefont {Linderberg}(1977)}]{beebe1977simplifications}%
  \BibitemOpen
  \bibfield  {author} {\bibinfo {author} {\bibfnamefont {N.~H.}\ \bibnamefont {Beebe}}\ and\ \bibinfo {author} {\bibfnamefont {J.}~\bibnamefont {Linderberg}},\ }\bibfield  {title} {\bibinfo {title} {Simplifications in the generation and transformation of two-electron integrals in molecular calculations},\ }\href {https://doi.org/10.1002/qua.560120408} {\bibfield  {journal} {\bibinfo  {journal} {Int. J. Quantum Chem}\ }\textbf {\bibinfo {volume} {12}},\ \bibinfo {pages} {683} (\bibinfo {year} {1977})}\BibitemShut {NoStop}%
\bibitem [{\citenamefont {Purwanto}\ \emph {et~al.}(2011)\citenamefont {Purwanto}, \citenamefont {Krakauer}, \citenamefont {Virgus},\ and\ \citenamefont {Zhang}}]{purwanto2011assessing}%
  \BibitemOpen
  \bibfield  {author} {\bibinfo {author} {\bibfnamefont {W.}~\bibnamefont {Purwanto}}, \bibinfo {author} {\bibfnamefont {H.}~\bibnamefont {Krakauer}}, \bibinfo {author} {\bibfnamefont {Y.}~\bibnamefont {Virgus}},\ and\ \bibinfo {author} {\bibfnamefont {S.}~\bibnamefont {Zhang}},\ }\bibfield  {title} {\bibinfo {title} {Assessing weak hydrogen binding on {Ca$^+$} centers: an accurate many-body study with large basis sets},\ }\href {https://doi.org/10.1063/1.3654002} {\bibfield  {journal} {\bibinfo  {journal} {J. Chem. Phys}\ }\textbf {\bibinfo {volume} {135}},\ \bibinfo {pages} {164105} (\bibinfo {year} {2011})}\BibitemShut {NoStop}%
\bibitem [{\citenamefont {Rubin}\ \emph {et~al.}(2018)\citenamefont {Rubin}, \citenamefont {Babbush},\ and\ \citenamefont {McClean}}]{rubin2018application}%
  \BibitemOpen
  \bibfield  {author} {\bibinfo {author} {\bibfnamefont {N.~C.}\ \bibnamefont {Rubin}}, \bibinfo {author} {\bibfnamefont {R.}~\bibnamefont {Babbush}},\ and\ \bibinfo {author} {\bibfnamefont {J.}~\bibnamefont {McClean}},\ }\bibfield  {title} {\bibinfo {title} {Application of fermionic marginal constraints to hybrid quantum algorithms},\ }\href {https://iopscience.iop.org/article/10.1088/1367-2630/aab919} {\bibfield  {journal} {\bibinfo  {journal} {New J. Phys}\ }\textbf {\bibinfo {volume} {20}},\ \bibinfo {pages} {053020} (\bibinfo {year} {2018})}\BibitemShut {NoStop}%
\bibitem [{\citenamefont {Motta}\ \emph {et~al.}(2019)\citenamefont {Motta}, \citenamefont {Shee}, \citenamefont {Zhang},\ and\ \citenamefont {Chan}}]{motta2019efficient}%
  \BibitemOpen
  \bibfield  {author} {\bibinfo {author} {\bibfnamefont {M.}~\bibnamefont {Motta}}, \bibinfo {author} {\bibfnamefont {J.}~\bibnamefont {Shee}}, \bibinfo {author} {\bibfnamefont {S.}~\bibnamefont {Zhang}},\ and\ \bibinfo {author} {\bibfnamefont {G.~K.-L.}\ \bibnamefont {Chan}},\ }\bibfield  {title} {\bibinfo {title} {Efficient ab initio auxiliary-field quantum {M}onte {C}arlo calculations in {G}aussian bases via low-rank tensor decomposition},\ }\href {https://pubs.acs.org/doi/abs/10.1021/acs.jctc.8b00996} {\bibfield  {journal} {\bibinfo  {journal} {J. Chem. Theory Comput}\ }\textbf {\bibinfo {volume} {15}},\ \bibinfo {pages} {3510} (\bibinfo {year} {2019})}\BibitemShut {NoStop}%
\bibitem [{\citenamefont {Berry}\ \emph {et~al.}(2019)\citenamefont {Berry}, \citenamefont {Gidney}, \citenamefont {Motta}, \citenamefont {McClean},\ and\ \citenamefont {Babbush}}]{berry2019qubitization}%
  \BibitemOpen
  \bibfield  {author} {\bibinfo {author} {\bibfnamefont {D.~W.}\ \bibnamefont {Berry}}, \bibinfo {author} {\bibfnamefont {C.}~\bibnamefont {Gidney}}, \bibinfo {author} {\bibfnamefont {M.}~\bibnamefont {Motta}}, \bibinfo {author} {\bibfnamefont {J.~R.}\ \bibnamefont {McClean}},\ and\ \bibinfo {author} {\bibfnamefont {R.}~\bibnamefont {Babbush}},\ }\bibfield  {title} {\bibinfo {title} {Qubitization of arbitrary basis quantum chemistry leveraging sparsity and low rank factorization},\ }\href {https://quantum-journal.org/papers/q-2019-12-02-208/} {\bibfield  {journal} {\bibinfo  {journal} {Quantum}\ }\textbf {\bibinfo {volume} {3}},\ \bibinfo {pages} {208} (\bibinfo {year} {2019})}\BibitemShut {NoStop}%
\bibitem [{\citenamefont {Lee}\ \emph {et~al.}(2021)\citenamefont {Lee}, \citenamefont {Berry}, \citenamefont {Gidney}, \citenamefont {Huggins}, \citenamefont {McClean}, \citenamefont {Wiebe},\ and\ \citenamefont {Babbush}}]{lee2021even}%
  \BibitemOpen
  \bibfield  {author} {\bibinfo {author} {\bibfnamefont {J.}~\bibnamefont {Lee}}, \bibinfo {author} {\bibfnamefont {D.~W.}\ \bibnamefont {Berry}}, \bibinfo {author} {\bibfnamefont {C.}~\bibnamefont {Gidney}}, \bibinfo {author} {\bibfnamefont {W.~J.}\ \bibnamefont {Huggins}}, \bibinfo {author} {\bibfnamefont {J.~R.}\ \bibnamefont {McClean}}, \bibinfo {author} {\bibfnamefont {N.}~\bibnamefont {Wiebe}},\ and\ \bibinfo {author} {\bibfnamefont {R.}~\bibnamefont {Babbush}},\ }\bibfield  {title} {\bibinfo {title} {Even more efficient quantum computations of chemistry through tensor hypercontraction},\ }\href {https://journals.aps.org/prxquantum/abstract/10.1103/PRXQuantum.2.030305} {\bibfield  {journal} {\bibinfo  {journal} {PRX Quantum}\ }\textbf {\bibinfo {volume} {2}},\ \bibinfo {pages} {030305} (\bibinfo {year} {2021})}\BibitemShut {NoStop}%
\bibitem [{\citenamefont {von Burg}\ \emph {et~al.}(2021)\citenamefont {von Burg}, \citenamefont {Low}, \citenamefont {H{\"a}ner}, \citenamefont {Steiger}, \citenamefont {Reiher}, \citenamefont {Roetteler},\ and\ \citenamefont {Troyer}}]{von2021quantum}%
  \BibitemOpen
  \bibfield  {author} {\bibinfo {author} {\bibfnamefont {V.}~\bibnamefont {von Burg}}, \bibinfo {author} {\bibfnamefont {G.~H.}\ \bibnamefont {Low}}, \bibinfo {author} {\bibfnamefont {T.}~\bibnamefont {H{\"a}ner}}, \bibinfo {author} {\bibfnamefont {D.~S.}\ \bibnamefont {Steiger}}, \bibinfo {author} {\bibfnamefont {M.}~\bibnamefont {Reiher}}, \bibinfo {author} {\bibfnamefont {M.}~\bibnamefont {Roetteler}},\ and\ \bibinfo {author} {\bibfnamefont {M.}~\bibnamefont {Troyer}},\ }\bibfield  {title} {\bibinfo {title} {Quantum computing enhanced computational catalysis},\ }\href {https://journals.aps.org/prresearch/abstract/10.1103/PhysRevResearch.3.033055} {\bibfield  {journal} {\bibinfo  {journal} {Phys. Rev. Research}\ }\textbf {\bibinfo {volume} {3}},\ \bibinfo {pages} {033055} (\bibinfo {year} {2021})}\BibitemShut {NoStop}%
\bibitem [{\citenamefont {Wierichs}\ \emph {et~al.}(2022)\citenamefont {Wierichs}, \citenamefont {Izaac}, \citenamefont {Wang},\ and\ \citenamefont {Lin}}]{wierichs2022general}%
  \BibitemOpen
  \bibfield  {author} {\bibinfo {author} {\bibfnamefont {D.}~\bibnamefont {Wierichs}}, \bibinfo {author} {\bibfnamefont {J.}~\bibnamefont {Izaac}}, \bibinfo {author} {\bibfnamefont {C.}~\bibnamefont {Wang}},\ and\ \bibinfo {author} {\bibfnamefont {C.~Y.-Y.}\ \bibnamefont {Lin}},\ }\bibfield  {title} {\bibinfo {title} {General parameter-shift rules for quantum gradients},\ }\href {https://quantum-journal.org/papers/q-2022-03-30-677/} {\bibfield  {journal} {\bibinfo  {journal} {Quantum}\ }\textbf {\bibinfo {volume} {6}},\ \bibinfo {pages} {677} (\bibinfo {year} {2022})}\BibitemShut {NoStop}%
\bibitem [{\citenamefont {Hubregtsen}\ \emph {et~al.}(2022)\citenamefont {Hubregtsen}, \citenamefont {Wilde}, \citenamefont {Qasim},\ and\ \citenamefont {Eisert}}]{hubregtsen2022single}%
  \BibitemOpen
  \bibfield  {author} {\bibinfo {author} {\bibfnamefont {T.}~\bibnamefont {Hubregtsen}}, \bibinfo {author} {\bibfnamefont {F.}~\bibnamefont {Wilde}}, \bibinfo {author} {\bibfnamefont {S.}~\bibnamefont {Qasim}},\ and\ \bibinfo {author} {\bibfnamefont {J.}~\bibnamefont {Eisert}},\ }\bibfield  {title} {\bibinfo {title} {Single-component gradient rules for variational quantum algorithms},\ }\href {https://iopscience.iop.org/article/10.1088/2058-9565/ac6824/meta} {\bibfield  {journal} {\bibinfo  {journal} {Quant. Sci. Tech}\ }\textbf {\bibinfo {volume} {7}},\ \bibinfo {pages} {035008} (\bibinfo {year} {2022})}\BibitemShut {NoStop}%
\bibitem [{\citenamefont {Izmaylov}\ \emph {et~al.}(2021)\citenamefont {Izmaylov}, \citenamefont {Lang},\ and\ \citenamefont {Yen}}]{izmaylov2021analytic}%
  \BibitemOpen
  \bibfield  {author} {\bibinfo {author} {\bibfnamefont {A.~F.}\ \bibnamefont {Izmaylov}}, \bibinfo {author} {\bibfnamefont {R.~A.}\ \bibnamefont {Lang}},\ and\ \bibinfo {author} {\bibfnamefont {T.-C.}\ \bibnamefont {Yen}},\ }\bibfield  {title} {\bibinfo {title} {Analytic gradients in variational quantum algorithms: Algebraic extensions of the parameter-shift rule to general unitary transformations},\ }\href@noop {} {\bibfield  {journal} {\bibinfo  {journal} {Physical Review A}\ }\textbf {\bibinfo {volume} {104}},\ \bibinfo {pages} {062443} (\bibinfo {year} {2021})}\BibitemShut {NoStop}%
\bibitem [{\citenamefont {Meyer}(2021)}]{meyer2021fisher}%
  \BibitemOpen
  \bibfield  {author} {\bibinfo {author} {\bibfnamefont {J.~J.}\ \bibnamefont {Meyer}},\ }\bibfield  {title} {\bibinfo {title} {Fisher information in noisy intermediate-scale quantum applications},\ }\href {https://quantum-journal.org/papers/q-2021-09-09-539/} {\bibfield  {journal} {\bibinfo  {journal} {Quantum}\ }\textbf {\bibinfo {volume} {5}},\ \bibinfo {pages} {539} (\bibinfo {year} {2021})}\BibitemShut {NoStop}%
\bibitem [{\citenamefont {Stokes}\ \emph {et~al.}(2020)\citenamefont {Stokes}, \citenamefont {Izaac}, \citenamefont {Killoran},\ and\ \citenamefont {Carleo}}]{stokes2020quantum}%
  \BibitemOpen
  \bibfield  {author} {\bibinfo {author} {\bibfnamefont {J.}~\bibnamefont {Stokes}}, \bibinfo {author} {\bibfnamefont {J.}~\bibnamefont {Izaac}}, \bibinfo {author} {\bibfnamefont {N.}~\bibnamefont {Killoran}},\ and\ \bibinfo {author} {\bibfnamefont {G.}~\bibnamefont {Carleo}},\ }\bibfield  {title} {\bibinfo {title} {Quantum natural gradient},\ }\href {https://quantum-journal.org/papers/q-2020-05-25-269/} {\bibfield  {journal} {\bibinfo  {journal} {Quantum}\ }\textbf {\bibinfo {volume} {4}},\ \bibinfo {pages} {269} (\bibinfo {year} {2020})}\BibitemShut {NoStop}%
\bibitem [{\citenamefont {Somma}\ \emph {et~al.}(2002)\citenamefont {Somma}, \citenamefont {Ortiz}, \citenamefont {Gubernatis}, \citenamefont {Knill},\ and\ \citenamefont {Laflamme}}]{somma2002simulating}%
  \BibitemOpen
  \bibfield  {author} {\bibinfo {author} {\bibfnamefont {R.}~\bibnamefont {Somma}}, \bibinfo {author} {\bibfnamefont {G.}~\bibnamefont {Ortiz}}, \bibinfo {author} {\bibfnamefont {J.~E.}\ \bibnamefont {Gubernatis}}, \bibinfo {author} {\bibfnamefont {E.}~\bibnamefont {Knill}},\ and\ \bibinfo {author} {\bibfnamefont {R.}~\bibnamefont {Laflamme}},\ }\bibfield  {title} {\bibinfo {title} {Simulating physical phenomena by quantum networks},\ }\href {https://journals.aps.org/pra/abstract/10.1103/PhysRevA.65.042323} {\bibfield  {journal} {\bibinfo  {journal} {Phys. Rev. A}\ }\textbf {\bibinfo {volume} {65}},\ \bibinfo {pages} {042323} (\bibinfo {year} {2002})}\BibitemShut {NoStop}%
\bibitem [{\citenamefont {Mitarai}\ and\ \citenamefont {Fujii}(2019)}]{mitarai2019methodology}%
  \BibitemOpen
  \bibfield  {author} {\bibinfo {author} {\bibfnamefont {K.}~\bibnamefont {Mitarai}}\ and\ \bibinfo {author} {\bibfnamefont {K.}~\bibnamefont {Fujii}},\ }\bibfield  {title} {\bibinfo {title} {Methodology for replacing indirect measurements with direct measurements},\ }\href {https://journals.aps.org/prresearch/abstract/10.1103/PhysRevResearch.1.013006} {\bibfield  {journal} {\bibinfo  {journal} {Phys. Rev. Research}\ }\textbf {\bibinfo {volume} {1}},\ \bibinfo {pages} {013006} (\bibinfo {year} {2019})}\BibitemShut {NoStop}%
\bibitem [{\citenamefont {Mitarai}\ and\ \citenamefont {Fujii}(2021{\natexlab{a}})}]{mitarai2021overhead}%
  \BibitemOpen
  \bibfield  {author} {\bibinfo {author} {\bibfnamefont {K.}~\bibnamefont {Mitarai}}\ and\ \bibinfo {author} {\bibfnamefont {K.}~\bibnamefont {Fujii}},\ }\bibfield  {title} {\bibinfo {title} {Overhead for simulating a non-local channel with local channels by quasiprobability sampling},\ }\href {https://quantum-journal.org/papers/q-2021-01-28-388/} {\bibfield  {journal} {\bibinfo  {journal} {Quantum}\ }\textbf {\bibinfo {volume} {5}},\ \bibinfo {pages} {388} (\bibinfo {year} {2021}{\natexlab{a}})}\BibitemShut {NoStop}%
\bibitem [{\citenamefont {Mitarai}\ and\ \citenamefont {Fujii}(2021{\natexlab{b}})}]{mitarai2021constructing}%
  \BibitemOpen
  \bibfield  {author} {\bibinfo {author} {\bibfnamefont {K.}~\bibnamefont {Mitarai}}\ and\ \bibinfo {author} {\bibfnamefont {K.}~\bibnamefont {Fujii}},\ }\bibfield  {title} {\bibinfo {title} {Constructing a virtual two-qubit gate by sampling single-qubit operations},\ }\href {https://iopscience.iop.org/article/10.1088/1367-2630/abd7bc} {\bibfield  {journal} {\bibinfo  {journal} {New J. Phys}\ }\textbf {\bibinfo {volume} {23}},\ \bibinfo {pages} {023021} (\bibinfo {year} {2021}{\natexlab{b}})}\BibitemShut {NoStop}%
\bibitem [{\citenamefont {Kinoshita}\ \emph {et~al.}(2005)\citenamefont {Kinoshita}, \citenamefont {Hino},\ and\ \citenamefont {Bartlett}}]{kinoshita2005coupled}%
  \BibitemOpen
  \bibfield  {author} {\bibinfo {author} {\bibfnamefont {T.}~\bibnamefont {Kinoshita}}, \bibinfo {author} {\bibfnamefont {O.}~\bibnamefont {Hino}},\ and\ \bibinfo {author} {\bibfnamefont {R.~J.}\ \bibnamefont {Bartlett}},\ }\bibfield  {title} {\bibinfo {title} {Coupled-cluster method tailored by configuration interaction},\ }\href {https://doi.org/10.1063/1.2000251} {\bibfield  {journal} {\bibinfo  {journal} {J. Chem. Phys}\ }\textbf {\bibinfo {volume} {123}},\ \bibinfo {pages} {074106} (\bibinfo {year} {2005})}\BibitemShut {NoStop}%
\bibitem [{\citenamefont {M{\"o}rchen}\ \emph {et~al.}(2020)\citenamefont {M{\"o}rchen}, \citenamefont {Freitag},\ and\ \citenamefont {Reiher}}]{morchen2020tailored}%
  \BibitemOpen
  \bibfield  {author} {\bibinfo {author} {\bibfnamefont {M.}~\bibnamefont {M{\"o}rchen}}, \bibinfo {author} {\bibfnamefont {L.}~\bibnamefont {Freitag}},\ and\ \bibinfo {author} {\bibfnamefont {M.}~\bibnamefont {Reiher}},\ }\bibfield  {title} {\bibinfo {title} {Tailored coupled cluster theory in varying correlation regimes},\ }\href@noop {} {\bibfield  {journal} {\bibinfo  {journal} {J. Chem. Phys}\ }\textbf {\bibinfo {volume} {153}},\ \bibinfo {pages} {244113} (\bibinfo {year} {2020})}\BibitemShut {NoStop}%
\bibitem [{\citenamefont {Faulstich}\ \emph {et~al.}(2019)\citenamefont {Faulstich}, \citenamefont {Laestadius}, \citenamefont {Legeza}, \citenamefont {Schneider},\ and\ \citenamefont {Kvaal}}]{faulstich2019analysis}%
  \BibitemOpen
  \bibfield  {author} {\bibinfo {author} {\bibfnamefont {F.~M.}\ \bibnamefont {Faulstich}}, \bibinfo {author} {\bibfnamefont {A.}~\bibnamefont {Laestadius}}, \bibinfo {author} {\bibfnamefont {O.}~\bibnamefont {Legeza}}, \bibinfo {author} {\bibfnamefont {R.}~\bibnamefont {Schneider}},\ and\ \bibinfo {author} {\bibfnamefont {S.}~\bibnamefont {Kvaal}},\ }\bibfield  {title} {\bibinfo {title} {Analysis of the tailored coupled-cluster method in quantum chemistry},\ }\href@noop {} {\bibfield  {journal} {\bibinfo  {journal} {SIAM Journal on Numerical Analysis}\ }\textbf {\bibinfo {volume} {57}},\ \bibinfo {pages} {2579} (\bibinfo {year} {2019})}\BibitemShut {NoStop}%
\bibitem [{\citenamefont {Scheurer}\ \emph {et~al.}(2024)\citenamefont {Scheurer}, \citenamefont {Anselmetti}, \citenamefont {Oumarou}, \citenamefont {Gogolin},\ and\ \citenamefont {Rubin}}]{scheurer2024tailored}%
  \BibitemOpen
  \bibfield  {author} {\bibinfo {author} {\bibfnamefont {M.}~\bibnamefont {Scheurer}}, \bibinfo {author} {\bibfnamefont {G.-L.~R.}\ \bibnamefont {Anselmetti}}, \bibinfo {author} {\bibfnamefont {O.}~\bibnamefont {Oumarou}}, \bibinfo {author} {\bibfnamefont {C.}~\bibnamefont {Gogolin}},\ and\ \bibinfo {author} {\bibfnamefont {N.~C.}\ \bibnamefont {Rubin}},\ }\bibfield  {title} {\bibinfo {title} {Tailored and externally corrected coupled cluster with quantum inputs},\ }\href@noop {} {\bibfield  {journal} {\bibinfo  {journal} {Journal of Chemical Theory and Computation}\ }\textbf {\bibinfo {volume} {20}},\ \bibinfo {pages} {5068} (\bibinfo {year} {2024})}\BibitemShut {NoStop}%
\bibitem [{\citenamefont {Robledo-Moreno}\ \emph {et~al.}(2024)\citenamefont {Robledo-Moreno}, \citenamefont {Motta}, \citenamefont {Haas}, \citenamefont {Javadi-Abhari}, \citenamefont {Jurcevic}, \citenamefont {Kirby}, \citenamefont {Martiel}, \citenamefont {Sharma}, \citenamefont {Sharma}, \citenamefont {Shirakawa} \emph {et~al.}}]{robledo2024chemistry}%
  \BibitemOpen
  \bibfield  {author} {\bibinfo {author} {\bibfnamefont {J.}~\bibnamefont {Robledo-Moreno}}, \bibinfo {author} {\bibfnamefont {M.}~\bibnamefont {Motta}}, \bibinfo {author} {\bibfnamefont {H.}~\bibnamefont {Haas}}, \bibinfo {author} {\bibfnamefont {A.}~\bibnamefont {Javadi-Abhari}}, \bibinfo {author} {\bibfnamefont {P.}~\bibnamefont {Jurcevic}}, \bibinfo {author} {\bibfnamefont {W.}~\bibnamefont {Kirby}}, \bibinfo {author} {\bibfnamefont {S.}~\bibnamefont {Martiel}}, \bibinfo {author} {\bibfnamefont {K.}~\bibnamefont {Sharma}}, \bibinfo {author} {\bibfnamefont {S.}~\bibnamefont {Sharma}}, \bibinfo {author} {\bibfnamefont {T.}~\bibnamefont {Shirakawa}}, \emph {et~al.},\ }\bibfield  {title} {\bibinfo {title} {Chemistry beyond exact solutions on a quantum-centric supercomputer},\ }\href@noop {} {\bibfield  {journal} {\bibinfo  {journal} {arXiv preprint arXiv:2405.05068}\ } (\bibinfo {year} {2024})}\BibitemShut {NoStop}%
\bibitem [{\citenamefont {Shi}\ \emph {et~al.}(2022)\citenamefont {Shi}, \citenamefont {Xie}, \citenamefont {Byrd},\ and\ \citenamefont {Nocedal}}]{shi2022noise}%
  \BibitemOpen
  \bibfield  {author} {\bibinfo {author} {\bibfnamefont {H.-J.~M.}\ \bibnamefont {Shi}}, \bibinfo {author} {\bibfnamefont {Y.}~\bibnamefont {Xie}}, \bibinfo {author} {\bibfnamefont {R.}~\bibnamefont {Byrd}},\ and\ \bibinfo {author} {\bibfnamefont {J.}~\bibnamefont {Nocedal}},\ }\bibfield  {title} {\bibinfo {title} {A noise-tolerant {quasi-Newton} algorithm for unconstrained optimization},\ }\href {https://dl.acm.org/doi/10.1137/20M1373190} {\bibfield  {journal} {\bibinfo  {journal} {SIAM J. Opt}\ }\textbf {\bibinfo {volume} {32}},\ \bibinfo {pages} {29} (\bibinfo {year} {2022})}\BibitemShut {NoStop}%
\bibitem [{\citenamefont {Garner}\ and\ \citenamefont {Neuscamman}(2023)}]{garner2023improving}%
  \BibitemOpen
  \bibfield  {author} {\bibinfo {author} {\bibfnamefont {S.~M.}\ \bibnamefont {Garner}}\ and\ \bibinfo {author} {\bibfnamefont {E.}~\bibnamefont {Neuscamman}},\ }\bibfield  {title} {\bibinfo {title} {Improving variational monte carlo optimization by avoiding statistically difficult parameters},\ }\href@noop {} {\bibfield  {journal} {\bibinfo  {journal} {arXiv preprint arXiv:2302.03078}\ } (\bibinfo {year} {2023})}\BibitemShut {NoStop}%
\bibitem [{\citenamefont {Arrasmith}\ \emph {et~al.}(2020)\citenamefont {Arrasmith}, \citenamefont {Cincio}, \citenamefont {Somma},\ and\ \citenamefont {Coles}}]{arrasmith2020operator}%
  \BibitemOpen
  \bibfield  {author} {\bibinfo {author} {\bibfnamefont {A.}~\bibnamefont {Arrasmith}}, \bibinfo {author} {\bibfnamefont {L.}~\bibnamefont {Cincio}}, \bibinfo {author} {\bibfnamefont {R.~D.}\ \bibnamefont {Somma}},\ and\ \bibinfo {author} {\bibfnamefont {P.~J.}\ \bibnamefont {Coles}},\ }\bibfield  {title} {\bibinfo {title} {Operator sampling for shot-frugal optimization in variational algorithms},\ }\href@noop {} {\bibfield  {journal} {\bibinfo  {journal} {arXiv:2004.06252}\ } (\bibinfo {year} {2020})}\BibitemShut {NoStop}%
\bibitem [{\citenamefont {Epperly}\ \emph {et~al.}(2022)\citenamefont {Epperly}, \citenamefont {Lin},\ and\ \citenamefont {Nakatsukasa}}]{epperly2022theory}%
  \BibitemOpen
  \bibfield  {author} {\bibinfo {author} {\bibfnamefont {E.~N.}\ \bibnamefont {Epperly}}, \bibinfo {author} {\bibfnamefont {L.}~\bibnamefont {Lin}},\ and\ \bibinfo {author} {\bibfnamefont {Y.}~\bibnamefont {Nakatsukasa}},\ }\bibfield  {title} {\bibinfo {title} {A theory of quantum subspace diagonalization},\ }\href@noop {} {\bibfield  {journal} {\bibinfo  {journal} {SIAM Journal on Matrix Analysis and Applications}\ }\textbf {\bibinfo {volume} {43}},\ \bibinfo {pages} {1263} (\bibinfo {year} {2022})}\BibitemShut {NoStop}%
\bibitem [{\citenamefont {Otis}\ and\ \citenamefont {Neuscamman}(2019)}]{otis2019complementary}%
  \BibitemOpen
  \bibfield  {author} {\bibinfo {author} {\bibfnamefont {L.}~\bibnamefont {Otis}}\ and\ \bibinfo {author} {\bibfnamefont {E.}~\bibnamefont {Neuscamman}},\ }\bibfield  {title} {\bibinfo {title} {Complementary first and second derivative methods for ansatz optimization in variational {Monte Carlo}},\ }\href {https://pubs.rsc.org/en/content/articlelanding/2019/cp/c9cp02269d} {\bibfield  {journal} {\bibinfo  {journal} {Phys. Chem. Chem. Phys}\ }\textbf {\bibinfo {volume} {21}},\ \bibinfo {pages} {14491} (\bibinfo {year} {2019})}\BibitemShut {NoStop}%
\bibitem [{\citenamefont {Zhao}\ and\ \citenamefont {Neuscamman}(2017)}]{zhao2017blocked}%
  \BibitemOpen
  \bibfield  {author} {\bibinfo {author} {\bibfnamefont {L.}~\bibnamefont {Zhao}}\ and\ \bibinfo {author} {\bibfnamefont {E.}~\bibnamefont {Neuscamman}},\ }\bibfield  {title} {\bibinfo {title} {A blocked linear method for optimizing large parameter sets in variational {Monte Carlo}},\ }\href {https://pubs.acs.org/doi/full/10.1021/acs.jctc.7b00119} {\bibfield  {journal} {\bibinfo  {journal} {J. Chem. Theory Comput}\ }\textbf {\bibinfo {volume} {13}},\ \bibinfo {pages} {2604} (\bibinfo {year} {2017})}\BibitemShut {NoStop}%
\bibitem [{\citenamefont {Neuscamman}\ \emph {et~al.}(2012)\citenamefont {Neuscamman}, \citenamefont {Umrigar},\ and\ \citenamefont {Chan}}]{neuscamman2012optimizing}%
  \BibitemOpen
  \bibfield  {author} {\bibinfo {author} {\bibfnamefont {E.}~\bibnamefont {Neuscamman}}, \bibinfo {author} {\bibfnamefont {C.}~\bibnamefont {Umrigar}},\ and\ \bibinfo {author} {\bibfnamefont {G.~K.-L.}\ \bibnamefont {Chan}},\ }\bibfield  {title} {\bibinfo {title} {Optimizing large parameter sets in variational quantum {Monte Carlo}},\ }\href {https://journals.aps.org/prb/abstract/10.1103/PhysRevB.85.045103} {\bibfield  {journal} {\bibinfo  {journal} {Phys. Rev. B}\ }\textbf {\bibinfo {volume} {85}},\ \bibinfo {pages} {045103} (\bibinfo {year} {2012})}\BibitemShut {NoStop}%
\bibitem [{\citenamefont {Scuseria}\ \emph {et~al.}(2011)\citenamefont {Scuseria}, \citenamefont {Jim{\'e}nez-Hoyos}, \citenamefont {Henderson}, \citenamefont {Samanta},\ and\ \citenamefont {Ellis}}]{scuseria2011projected}%
  \BibitemOpen
  \bibfield  {author} {\bibinfo {author} {\bibfnamefont {G.~E.}\ \bibnamefont {Scuseria}}, \bibinfo {author} {\bibfnamefont {C.~A.}\ \bibnamefont {Jim{\'e}nez-Hoyos}}, \bibinfo {author} {\bibfnamefont {T.~M.}\ \bibnamefont {Henderson}}, \bibinfo {author} {\bibfnamefont {K.}~\bibnamefont {Samanta}},\ and\ \bibinfo {author} {\bibfnamefont {J.~K.}\ \bibnamefont {Ellis}},\ }\bibfield  {title} {\bibinfo {title} {Projected quasiparticle theory for molecular electronic structure},\ }\href {https://aip.scitation.org/doi/10.1063/1.3643338} {\bibfield  {journal} {\bibinfo  {journal} {J. Chem. Phys}\ }\textbf {\bibinfo {volume} {135}},\ \bibinfo {pages} {124108} (\bibinfo {year} {2011})}\BibitemShut {NoStop}%
\bibitem [{\citenamefont {Jim{\'e}nez-Hoyos}\ \emph {et~al.}(2012)\citenamefont {Jim{\'e}nez-Hoyos}, \citenamefont {Henderson}, \citenamefont {Tsuchimochi},\ and\ \citenamefont {Scuseria}}]{jimenez2012projected}%
  \BibitemOpen
  \bibfield  {author} {\bibinfo {author} {\bibfnamefont {C.~A.}\ \bibnamefont {Jim{\'e}nez-Hoyos}}, \bibinfo {author} {\bibfnamefont {T.~M.}\ \bibnamefont {Henderson}}, \bibinfo {author} {\bibfnamefont {T.}~\bibnamefont {Tsuchimochi}},\ and\ \bibinfo {author} {\bibfnamefont {G.~E.}\ \bibnamefont {Scuseria}},\ }\bibfield  {title} {\bibinfo {title} {Projected {Hartree--Fock} theory},\ }\href {https://aip.scitation.org/doi/10.1063/1.4705280} {\bibfield  {journal} {\bibinfo  {journal} {J. Chem. Phys}\ }\textbf {\bibinfo {volume} {136}},\ \bibinfo {pages} {164109} (\bibinfo {year} {2012})}\BibitemShut {NoStop}%
\bibitem [{\citenamefont {Shi}\ \emph {et~al.}(2014)\citenamefont {Shi}, \citenamefont {Jim{\'e}nez-Hoyos}, \citenamefont {Rodr{\'\i}guez-Guzm{\'a}n}, \citenamefont {Scuseria},\ and\ \citenamefont {Zhang}}]{shi2014symmetry}%
  \BibitemOpen
  \bibfield  {author} {\bibinfo {author} {\bibfnamefont {H.}~\bibnamefont {Shi}}, \bibinfo {author} {\bibfnamefont {C.~A.}\ \bibnamefont {Jim{\'e}nez-Hoyos}}, \bibinfo {author} {\bibfnamefont {R.}~\bibnamefont {Rodr{\'\i}guez-Guzm{\'a}n}}, \bibinfo {author} {\bibfnamefont {G.~E.}\ \bibnamefont {Scuseria}},\ and\ \bibinfo {author} {\bibfnamefont {S.}~\bibnamefont {Zhang}},\ }\bibfield  {title} {\bibinfo {title} {Symmetry-projected wave functions in quantum {Monte Carlo} calculations},\ }\href {https://journals.aps.org/prb/abstract/10.1103/PhysRevB.89.125129} {\bibfield  {journal} {\bibinfo  {journal} {Phys. Rev. B}\ }\textbf {\bibinfo {volume} {89}},\ \bibinfo {pages} {125129} (\bibinfo {year} {2014})}\BibitemShut {NoStop}%
\bibitem [{\citenamefont {Qiu}\ \emph {et~al.}(2017)\citenamefont {Qiu}, \citenamefont {Henderson}, \citenamefont {Zhao},\ and\ \citenamefont {Scuseria}}]{qiu2017projected}%
  \BibitemOpen
  \bibfield  {author} {\bibinfo {author} {\bibfnamefont {Y.}~\bibnamefont {Qiu}}, \bibinfo {author} {\bibfnamefont {T.~M.}\ \bibnamefont {Henderson}}, \bibinfo {author} {\bibfnamefont {J.}~\bibnamefont {Zhao}},\ and\ \bibinfo {author} {\bibfnamefont {G.~E.}\ \bibnamefont {Scuseria}},\ }\bibfield  {title} {\bibinfo {title} {Projected coupled cluster theory},\ }\href {https://doi.org/10.1063/1.4991020} {\bibfield  {journal} {\bibinfo  {journal} {J. Chem. Phys}\ }\textbf {\bibinfo {volume} {147}},\ \bibinfo {pages} {064111} (\bibinfo {year} {2017})}\BibitemShut {NoStop}%
\bibitem [{\citenamefont {Qiu}\ \emph {et~al.}(2018)\citenamefont {Qiu}, \citenamefont {Henderson}, \citenamefont {Zhao},\ and\ \citenamefont {Scuseria}}]{qiu2018projected}%
  \BibitemOpen
  \bibfield  {author} {\bibinfo {author} {\bibfnamefont {Y.}~\bibnamefont {Qiu}}, \bibinfo {author} {\bibfnamefont {T.~M.}\ \bibnamefont {Henderson}}, \bibinfo {author} {\bibfnamefont {J.}~\bibnamefont {Zhao}},\ and\ \bibinfo {author} {\bibfnamefont {G.~E.}\ \bibnamefont {Scuseria}},\ }\bibfield  {title} {\bibinfo {title} {Projected coupled cluster theory: optimization of cluster amplitudes in the presence of symmetry projection},\ }\href {https://doi.org/10.1063/1.5053605} {\bibfield  {journal} {\bibinfo  {journal} {J. Chem. Phys}\ }\textbf {\bibinfo {volume} {149}},\ \bibinfo {pages} {164108} (\bibinfo {year} {2018})}\BibitemShut {NoStop}%
\bibitem [{\citenamefont {Papastathopoulos-Katsaros}\ \emph {et~al.}(2023)\citenamefont {Papastathopoulos-Katsaros}, \citenamefont {Henderson},\ and\ \citenamefont {Scuseria}}]{papastathopoulos2023symmetry}%
  \BibitemOpen
  \bibfield  {author} {\bibinfo {author} {\bibfnamefont {A.}~\bibnamefont {Papastathopoulos-Katsaros}}, \bibinfo {author} {\bibfnamefont {T.~M.}\ \bibnamefont {Henderson}},\ and\ \bibinfo {author} {\bibfnamefont {G.~E.}\ \bibnamefont {Scuseria}},\ }\bibfield  {title} {\bibinfo {title} {Symmetry-projected cluster mean-field theory applied to spin systems},\ }\href {https://doi.org/10.1063/5.0155765} {\bibfield  {journal} {\bibinfo  {journal} {J. Chem. Phys}\ }\textbf {\bibinfo {volume} {159}},\ \bibinfo {pages} {084107} (\bibinfo {year} {2023})}\BibitemShut {NoStop}%
\bibitem [{\citenamefont {Yen}\ \emph {et~al.}(2019)\citenamefont {Yen}, \citenamefont {Lang},\ and\ \citenamefont {Izmaylov}}]{yen2019exact}%
  \BibitemOpen
  \bibfield  {author} {\bibinfo {author} {\bibfnamefont {T.-C.}\ \bibnamefont {Yen}}, \bibinfo {author} {\bibfnamefont {R.~A.}\ \bibnamefont {Lang}},\ and\ \bibinfo {author} {\bibfnamefont {A.~F.}\ \bibnamefont {Izmaylov}},\ }\bibfield  {title} {\bibinfo {title} {Exact and approximate symmetry projectors for the electronic structure problem on a quantum computer},\ }\href@noop {} {\bibfield  {journal} {\bibinfo  {journal} {The Journal of chemical physics}\ }\textbf {\bibinfo {volume} {151}} (\bibinfo {year} {2019})}\BibitemShut {NoStop}%
\bibitem [{\citenamefont {Bravyi}\ \emph {et~al.}(2017)\citenamefont {Bravyi}, \citenamefont {Gambetta}, \citenamefont {Mezzacapo},\ and\ \citenamefont {Temme}}]{bravyi2017tapering}%
  \BibitemOpen
  \bibfield  {author} {\bibinfo {author} {\bibfnamefont {S.}~\bibnamefont {Bravyi}}, \bibinfo {author} {\bibfnamefont {J.~M.}\ \bibnamefont {Gambetta}}, \bibinfo {author} {\bibfnamefont {A.}~\bibnamefont {Mezzacapo}},\ and\ \bibinfo {author} {\bibfnamefont {K.}~\bibnamefont {Temme}},\ }\bibfield  {title} {\bibinfo {title} {Tapering off qubits to simulate fermionic {Hamiltonians}},\ }\href {https://arxiv.org/abs/1701.08213} {\bibfield  {journal} {\bibinfo  {journal} {arXiv:1701.08213}\ } (\bibinfo {year} {2017})}\BibitemShut {NoStop}%
\bibitem [{\citenamefont {Gottesman}(2010)}]{gottesman2010introduction}%
  \BibitemOpen
  \bibfield  {author} {\bibinfo {author} {\bibfnamefont {D.}~\bibnamefont {Gottesman}},\ }\bibfield  {title} {\bibinfo {title} {An introduction to quantum error correction and fault-tolerant quantum computation},\ }\href {https://arxiv.org/abs/0904.2557} {\bibfield  {journal} {\bibinfo  {journal} {Proc. Symposia Appl. Math}\ }\textbf {\bibinfo {volume} {68}},\ \bibinfo {pages} {13} (\bibinfo {year} {2010})}\BibitemShut {NoStop}%
\bibitem [{\citenamefont {Shor}(1995)}]{shor1995scheme}%
  \BibitemOpen
  \bibfield  {author} {\bibinfo {author} {\bibfnamefont {P.~W.}\ \bibnamefont {Shor}},\ }\bibfield  {title} {\bibinfo {title} {Scheme for reducing decoherence in quantum computer memory},\ }\href {https://journals.aps.org/pra/abstract/10.1103/PhysRevA.52.R2493} {\bibfield  {journal} {\bibinfo  {journal} {Phys. Rev. A}\ }\textbf {\bibinfo {volume} {52}},\ \bibinfo {pages} {R2493} (\bibinfo {year} {1995})}\BibitemShut {NoStop}%
\bibitem [{\citenamefont {Aharonov}\ and\ \citenamefont {Ben-Or}(1997)}]{aharonov1997fault}%
  \BibitemOpen
  \bibfield  {author} {\bibinfo {author} {\bibfnamefont {D.}~\bibnamefont {Aharonov}}\ and\ \bibinfo {author} {\bibfnamefont {M.}~\bibnamefont {Ben-Or}},\ }\bibfield  {title} {\bibinfo {title} {Fault-tolerant quantum computation with constant error},\ }\href {https://dl.acm.org/doi/10.1145/258533.258579} {\bibfield  {journal} {\bibinfo  {journal} {Proc. ACM}\ ,\ \bibinfo {pages} {176}} (\bibinfo {year} {1997})}\BibitemShut {NoStop}%
\bibitem [{\citenamefont {Knill}\ \emph {et~al.}(1998)\citenamefont {Knill}, \citenamefont {Laflamme},\ and\ \citenamefont {Zurek}}]{knill1998resilient}%
  \BibitemOpen
  \bibfield  {author} {\bibinfo {author} {\bibfnamefont {E.}~\bibnamefont {Knill}}, \bibinfo {author} {\bibfnamefont {R.}~\bibnamefont {Laflamme}},\ and\ \bibinfo {author} {\bibfnamefont {W.~H.}\ \bibnamefont {Zurek}},\ }\bibfield  {title} {\bibinfo {title} {Resilient quantum computation},\ }\href {https://www.science.org/doi/10.1126/science.279.5349.342} {\bibfield  {journal} {\bibinfo  {journal} {Science}\ }\textbf {\bibinfo {volume} {279}},\ \bibinfo {pages} {342} (\bibinfo {year} {1998})}\BibitemShut {NoStop}%
\bibitem [{\citenamefont {Kitaev}(2003)}]{kitaev2003fault}%
  \BibitemOpen
  \bibfield  {author} {\bibinfo {author} {\bibfnamefont {A.~Y.}\ \bibnamefont {Kitaev}},\ }\bibfield  {title} {\bibinfo {title} {Fault-tolerant quantum computation by anyons},\ }\href {https://doi.org/10.1016/S0003-4916%2802%2900018-0} {\bibfield  {journal} {\bibinfo  {journal} {Ann. Phys}\ }\textbf {\bibinfo {volume} {303}},\ \bibinfo {pages} {2} (\bibinfo {year} {2003})}\BibitemShut {NoStop}%
\end{thebibliography}%

\clearpage 
\begin{figure}[h!]
\includegraphics[width=\textwidth]{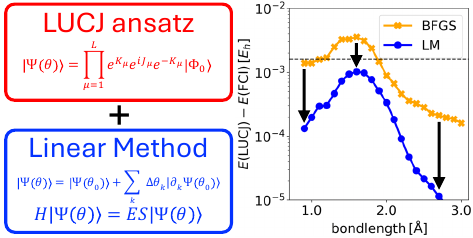}
\caption{TOC Graphic}
\end{figure}

\end{document}